\documentclass[aps,pre,superscriptaddress,twocolumn]{revtex4-2}

\preprint{Submitted to Phys. Rev. E}
\usepackage[T1]{fontenc}
\usepackage[utf8]{inputenc}
\usepackage{babel}

\usepackage{amsmath,amssymb,amsfonts}
\usepackage{bm}              
\usepackage{mathtools}       
\usepackage{physics}         

\usepackage{graphicx}
\usepackage{color}
\usepackage{float}
\usepackage{url}

\usepackage{booktabs}
\usepackage{array}

\usepackage{siunitx}
\usepackage[colorlinks=true,
            linkcolor=blue,
            citecolor=blue,
            urlcolor=blue]{hyperref}

\begin{document}

\title{Non-equilibrium (thermo)dynamics of colloids under mobile piston compression}

\author{Arturo Moncho-Jordá}
\email{moncho@ugr.es}
\thanks{Corresponding author}
\affiliation{Department of Applied Physics, University of Granada, 18071 Granada, Spain}
\affiliation{Institute
Carlos I for Theoretical and Computational Physics, University de Granada, 18071 Granada, Spain}
\author{José López-Molina}
\affiliation{Department of Health Sciences and Technology, ETH Zürich, Zürich 8092, Switzerland}

\author{Joachim Dzubiella}
\email{joachim.dzubiella@physik.uni-freiburg.de}
\thanks{Corresponding author}
\affiliation{Physikalisches Institut, Albert-Ludwigs- Universität Freiburg, D-79104 Freiburg, Germany}


\date{\today}

\begin{abstract}

We investigate the non-equilibrium compression of a confined colloidal fluid driven by a mobile boundary within dynamical density functional theory. The system consist of a hard-sphere fluid confined between two parallel walls, one of which acts as an overdamped piston subjected to a sudden increase in external pressure. The piston motion is characterized by a mobility parameter $K$. By varying $K$ over several orders of magnitude, we identify a crossover from quasi-static compression to a diffusion-limited strongly driven regime. For small $K$, the system evolves through near-equilibrium states and the total injected work attains its minimal value, equal to the equilibrium free-energy difference. In contrast, for large $K$, the piston rapidly adjusts and the dynamics becomes controlled by the intrinsic diffusive relaxation of the confined colloidal fluid, leading to universal saturation behavior of the piston trajectory, pressure--position relation, particle currents, and center-of-mass velocity. In this regime, the total injected work and entropy production are bounded, reflecting fundamental constraints imposed by diffusive transport. We find that the maximum injected power scales linearly with $K$, while the entropy-production peak exhibits a crossover from quadratic growth to saturation. The corresponding peak times display distinct $1/K$ asymptotic regimes separated by an intermediate crossover. The entropy change of the thermal bath is computed explicitly and shown to interpolate between the reversible limit, where it exactly compensates the configurational entropy loss of the fluid, and a strongly driven regime dominated by irreversible dissipation. Finally, the time evolution of the configurational entropy and the external potential energy reveals a dynamical decoupling between geometric confinement and structural relaxation, including transient non-monotonic behavior in the high-mobility regime. These results provide a quantitative thermodynamic characterization of boundary-driven compression and uncover generic non-equilibrium features governed by a single mobility parameter.
\end{abstract}

\maketitle

\noindent\textit{Submitted to Physical Review E}

\section{Introduction}

Understanding the non-equilibrium response of interacting many-particle systems to external driving is a central problem in soft-matter physics and statistical mechanics, see, for example~\cite{ Kubo:Science,wagner1989,Lowen_2001,seifert2012stochastic,marconi2012dynamic,Dzubiella_2003,Speck_2007, Brader30032011, PhysRevE.82.061503,PhysRevE.88.052307,Falasco2016, PhysRevLett.125.018001, PhysRevLett.121.098002,doi:10.1073/pnas.2510654122, Roichman:Nature,lopez-molina2024nonequilibrium,Straube_2024}. Colloidal suspensions provide a particularly well-suited model system in this context: their dynamics is intrinsically overdamped, thermal fluctuations play a dominant role, and interparticle interactions are tuneable and can often be described accurately by effective potentials. As a result, colloids offer a unique platform to investigate how microscopic interactions and constraints translate into macroscopic transport, relaxation, and dissipation under non-equilibrium conditions.

Confinement introduces an additional layer of complexity to the non-equilibrium dynamics of colloidal fluids~\cite{Lowen_2001}. When colloidal particles are restricted to narrow geometries, excluded-volume interactions lead to strong spatial inhomogeneities, such as density layering near confining boundaries, while transport and relaxation become highly anisotropic and history dependent~\cite{evans1979nature,roth2010fundamental,morais2024,Lowen2009TwentyYearsConfinedColloids,Schilling:review,Straube_2024,lopez-molina2024nonequilibrium}. External driving applied through the boundaries can then generate complex relaxation pathways, nontrivial current patterns, and irreversible dissipation that are absent in bulk systems. Understanding how confinement and boundary-driven forcing jointly control non-equilibrium behavior remains an open challenge.

Dynamic Density Functional Theory (DDFT) has been established as a powerful theoretical framework to describe the time evolution of interacting Brownian particles under time-dependent external fields~\cite{marconi1999dynamic,archer2004,espanol2009}, although based on approximations~\cite{vrugt2020,vrugt2022,Schmidt2022}. DDFT provides a microscopically motivated yet computationally tractable description of non-equilibrium dynamics, while retaining direct access to thermodynamic quantities such as free energy, work, and entropy production~\cite{wittkowski2012, Sekimoto2010, hatano2001steady}. When supplemented with accurate equilibrium free-energy functionals, DDFT yields a reliable description of the dynamics of dense fluids in strongly inhomogeneous environments.~\cite{vrugt2020}

Mechanical driving through time-dependent boundaries represents a particularly relevant and experimentally accessible class of non-equilibrium protocols. Examples include compression and expansion processes in microfluidic devices, confined colloidal assemblies, granular systems, and soft or active materials driven by movable walls or pistons~\cite{kestemont2000,brito2005,hurtado2006,leonard2013,bechinger2016active,Lopez2015}. In such situations, the boundary motion is not prescribed \textit{a priori} but emerges from the coupling between the externally imposed load and the instantaneous response of the confined fluid. In fact, in experiments moveable walls and pistons are not passive but typically a dynamic part of the total non-equilibrium system, e.g., in the uniaxial compression of colloids~\cite{piston_clay,yael:piston,piston_karg1,piston_karg2,Isa_compression,uniaxial_compression}. However, the specific action of the moveable boundaries on colloidal dynamics has not been systematically investigated so far. 

In this work, we investigate the full non-equilibrium, uniaxial compression of a hard-sphere fluid confined between two parallel walls, one of which acts as an overdamped mobile piston. The piston is permeable to the solvent: It acts only as a semipermeable piston with external osmotic pressure $P_\mathrm{ext}$. The aims of the study are manifold: i) introduce the methodology of a moving boundary into DDFT, ii) understand its action on the time-dependent structural evolution of the colloid, and iii) connect the latter to the resulting non-equilibrium thermodynamics in terms of entropy production, dissipation, and work.  For this, the piston dynamics is assumed to be overdamped and relates its velocity to the imbalance between the pressure exerted by the fluid and an externally imposed load. This setup naturally introduces a single control parameter—the piston mobility (or interpretable as inverse friction), $K$, which sets the relative timescale between mechanical driving and Brownian relaxation of the fluid.  By varying $K$ over several orders of magnitude, we are able to explore a broad range of dynamical regimes, from quasi-static compression to strongly driven, far-from-equilibrium dynamics.

Starting from an initial equilibrium, we consider a sudden increase of the piston pressure and analyze the relaxation process toward the final compressed equilibrium. Within the employed DDFT--FMT (Fundamental Measure Theory~\cite{hansengoos2006density}) framework, we track the time evolution of microscopic observables such as density profiles and particle currents, as well as macroscopic quantities including piston position, pressure buildup, mean fluid velocity, injected mechanical power, and entropy production. This allows us to establish a direct and quantitative connection between microscopic density rearrangements and macroscopic energetic balances.

Our results reveal distinct dynamical regimes controlled by $K$, including a quasi-static regime in which the system remains close to equilibrium, an intermediate regime characterized by strong non-equilibrium effects and pronounced spatial asymmetries, and a high-mobility regime in which the piston rapidly adjusts, and the relaxation becomes limited by the intrinsic diffusion of the fluid. In this case, we observe universal saturation behavior in the piston motion, pressure response, particle currents, and non-equilibrium thermodynamic quantities. In particular, both the injected work and the total entropy production are shown to be bounded from above, reflecting fundamental limitations imposed by diffusive transport and confinement.


The paper is organized as follows. In Sec.~\ref{sec:theory} we present the theoretical framework, including the DDFT formalism, the coupling to an overdamped mobile piston, and the identification of work and entropy production. Sec.~\ref{sec:numerical} describes the numerical implementation and the observables used to characterize the dynamics. In Sec.~\ref{sec:results} we present and discuss the results, highlighting the emergence of distinct dynamical regimes and universal scaling behaviors controlled by the piston mobility. Finally, Sec.~\ref{sec:conclusions} summarizes our findings and outlines possible extensions of the present approach.

\section{Theory}
\label{sec:theory}
\subsection{Dynamic Density Functional Theory}

We consider a colloidal system formed by $N$ hard spheres of radius $R$ subjected to a spatial and time dependent external potential $V_\mathrm{ext}(\mathbf{r},t)$. The non-equilibrium time evolution of the one-body density field $\rho(\mathbf r,t)$ is described within the framework of DDFT~\cite{marconi1999dynamic,archer2004,espanol2009, Archer2009}. The time evolution of the density follows a continuity equation,
\begin{equation}
\frac{\partial \rho(\mathbf r,t)}{\partial t}
= - \nabla \cdot \mathbf J(\mathbf r,t),
\label{eq:continuity}
\end{equation}
where $\mathbf J(\mathbf r,t)$ is the particle current. For overdamped Brownian dynamics, the current is given by
\begin{equation}
\mathbf J(\mathbf r,t)= - D \rho(\mathbf r,t)\,\nabla (\beta \mu(\mathbf r,t)),
\label{eq:current}
\end{equation}
where $\beta = (k_B T)^{-1}$, $D$ is the diffusion coefficient of the particles, and $\mu(\mathbf{r},t)$ is the non-equilibrium local chemical potential, defined as functional derivative of the total free energy functional, i.e. $\mu = \delta \mathcal{F}[\rho]/\delta\rho$.

The total free-energy functional entering DDFT is identical to that of equilibrium density functional theory,
\begin{equation}
\mathcal{F}[\rho] = F[\rho] + \int d\mathbf r\, \rho(\mathbf r,t) V_{\mathrm{ext}}(\mathbf r,t).
\label{eq:totalfreeenergy}
\end{equation}
where $F$ is the intrinsic free energy of the fluid. $F$ splits into ideal and excess contributions, $F[\rho]=F_{\mathrm{id}}[\rho] + F_{\mathrm{ex}}[\rho]$. The ideal-gas contribution reads
\begin{equation}
\beta F_{\mathrm{id}}[\rho]
= \int d\mathbf r\, \rho(\mathbf r,t)
\left[
\ln\!\left(\rho(\mathbf r,t)\Lambda^3\right) - 1
\right],
\end{equation}
where $\Lambda$ is the thermal de Broglie wavelength, while excess free energy $F_{\mathrm{ex}}[\rho]$ accounts for
interparticle interactions. The local chemical potential is given by
\begin{equation}
\beta\mu(\mathbf{r},t)=\ln\!\left(\rho(\mathbf r,t)\Lambda^3\right)+\frac{\delta \beta F_\mathrm{ex}[\rho]}{\delta \rho(\mathbf{r},t)}+\beta V_{\mathrm{ext}}(\mathbf{r},t).
\label{eq:chemical_potential}
\end{equation}

Combining Eqs.~\eqref{eq:continuity}, \eqref{eq:current} and \eqref{eq:chemical_potential} yields the DDFT equation
\begin{equation}
\frac{\partial \rho}{\partial t}=D\nabla\cdot\left[\nabla\rho +\rho\nabla \Big( \frac{\delta \beta F_\mathrm{ex}[\rho]}{\delta \rho}+\beta V_{\mathrm{ext}}(\mathbf{r},t)\Big) \right],
\label{eq:mainDDFT}
\end{equation}
yielding the time evolution of the particle number density, $\rho(\mathbf{r},t)$. It is important to emphasize that DDFT relies on the adiabatic approximation, whereby the non-equilibrium two-body correlations are assumed to be identical to those of an equilibrium system with the same instantaneous density profile~\cite{Archer2009,Schmidt2022}. Hydrodynamic interactions and inertial effects are neglected~\cite{rauscher2007dynamic, donev2014dynamic,goddard2016dynamical}, which is appropriate for overdamped colloidal suspensions at low Reynolds numbers.

The remaining ingredient required to close Eq.~\eqref{eq:mainDDFT} is a prescription for the inhomogeneous excess free-energy functional of a hard-sphere fluid. Our system is accurately described by the White Bear version (Mark II) of the FMT developed by Hansen-Goos and Roth~\cite{hansengoos2006density}.

\subsection{Coupled DDFT-overdamped mobile piston}

Our system is confined in the $z$-direction by two repulsive parallel walls of area $A$. The left wall is fixed at $z=0$, while the right wall is a mobile piston located at $z=L(t)$. Due to finite size of the hard colloids, the accessible region is $R \le z \le L(t)-R$. Given the symmetry of the system and imposing homogeneity in $xy$, the density profile is $\rho(\mathbf{r},t)=\rho(z,t)$. In addition to an impenetrable walls, both boundaries are modeled by short-ranged repulsive continuous  potentials (explicitly given later in section~\ref{sec:numerical}),
\begin{equation}
V_{\mathrm{ext}}(z,t)=V_\mathrm{L}(z) + V_\mathrm{R}(z,t),
\label{eq:Vext_soft}
\end{equation}
where $V_{\mathrm{L}}(z)$ represents the fixed
left wall, while $V_{\mathrm{R}}(z,t)=V_{\mathrm{R}}(z,L(t))$ accounts for the interaction with the mobile piston located at position $L(t)$. The explicit time dependence of the external potential arises solely from the piston motion.

The instantaneous pressure exerted by the colloidal fluid on the left wall ($P_\mathrm{L}$) and the piston ($P_\mathrm{R}$) follow directly from the wall
interactions, i.e.,
\begin{align}
\label{eq:PL}
P_\mathrm{L}=&-\int_R^{L(t)-R} dz\rho(z,t)\frac{\partial V_\mathrm{L}(z)}{\partial z}, \\
P_\mathrm{R}(t)=&\int_R^{L(t)-R} dz\rho(z,t)\frac{\partial V_\mathrm{R}(z,t)}{\partial z}.
\label{eq:PR}
\end{align}


The piston is assumed overdamped and evolves according to a linear mobility law,
\begin{equation}
\frac{dL(t)}{dt}=K\,\bigl(P_\mathrm{R}(t) - P_{\mathrm{ext}}\bigr),
\label{eq:piston_overdamped}
\end{equation}
where $P_{\mathrm{ext}}$ is the externally imposed pressure (load) acting on the piston and $K$ is the piston mobility coefficient, defined as $K=A/\xi$, where $A$ is the cross-sectional area of the piston and $\xi$ its intrinsic friction coefficient. $K$ sets the characteristic timescale of the mechanical response of the piston to pressure imbalances. In the limit of large $K$, the piston responds rapidly, leading to a strong coupling between the fluid dynamics and the boundary motion, whereas for small $K$ the piston dynamics is slow and effectively quasi-static from the perspective of the fluid. As a result, $K$ governs the degree of non-equilibrium during the compression process and plays a central role in controlling the colloidal relaxation pathways. In other words, the ratio between the piston timescale and that of the intrinsic diffusive relaxation of the liquid  determines whether the system evolves close to local equilibrium or is driven far from equilibrium during compression. 

Coupled Eqs.~\eqref{eq:mainDDFT}, \eqref{eq:PR}, and \eqref{eq:piston_overdamped} define a closed DDFT–piston problem. Starting from an initial density profile $\rho(z,0)$, an initial piston position $L_0$, and a piston mobility $K$, the density field and piston position are evolved self-consistently in time. In our protocol, the system is initially prepared in an equilibrium state under an applied piston pressure $P_0$. At $t=0$, the piston pressure is abruptly switched to a new value $P_{\mathrm{ext}}$, thereby inducing a compression of the fluid. The system subsequently relaxes toward a new equilibrium state consistent with the imposed pressure, $P_{\infty}=P_{\mathrm{ext}}$, at which the piston reaches its final position $L_{\infty}$.

\subsection{Non-equilibrium thermodynamics: identification of dissipation and work}

Monitoring the evolution of the fluid non-equilibrium free energy provides key insight into
the thermodynamic mechanisms underlying the compression dynamics. While applications of DDFT typically only focus on the time evolution of the density field, the free energy offers a complementary, physically transparent description that allows one to quantify the departure from equilibrium, the
directionality of the relaxation process in term of irreversible dissipation, and the energetic cost (work) associated with the piston-driven confinement. In particular, we can establish direct connections between microscopic density rearrangements and macroscopic mechanical work performed on the system. Note again that we neglect superadiabatic contributions to the dissipation~\cite{Schmidt2022}. 

The time derivative of the non-equilibrium free energy functional \(\mathcal{F}[\rho,t]\) is obtained by functional differentiation (c.f. Eq.~\eqref{eq:totalfreeenergy}):
\begin{align}
\frac{d\mathcal F[\rho;t]}{dt}
&= \int d\mathbf r\,
\left[
\frac{\delta F[\rho;t]}{\delta \rho(\mathbf r,t)}
+ V_{\mathrm{ext}}(\mathbf r,t)
\right]
\frac{\partial \rho(\mathbf r,t)}{\partial t}
\nonumber\\
&\quad
+ \int d\mathbf r\, \rho(\mathbf r,t)\,
\frac{\partial V_{\mathrm{ext}}(\mathbf r,t)}{\partial t}.
\label{eq:dFdt1}
\end{align}

Using Eq.~\eqref{eq:chemical_potential} we find that
\begin{equation}
\frac{d\mathcal F}{dt}= \int d\mathbf r\,\mu(\mathbf{r},t)\frac{\partial \rho(\mathbf r,t)}{\partial t}+\int d\mathbf r\, \rho(\mathbf r,t)\,
\frac{\partial V_{\mathrm{ext}}(\mathbf r,t)}{\partial t}.
\label{eq:dFdt2}
\end{equation}
 Focusing first on the former term, particle conservation in DDFT, expressed through $\partial_t\rho = -\nabla\cdot\mathbf J$, allows this contribution in Eq.~\eqref{eq:dFdt2} to be rewritten as
\begin{align}
\int d\mathbf r\, \mu\,\partial_t\rho
&=-\int d\mathbf r\, \mu(\mathbf r,t)\,\nabla\cdot \mathbf J(\mathbf r,t)\nonumber
\\
&=
-\int d\mathbf r\, \nabla\cdot\bigl(\mu\,\mathbf J\bigr)
+\int d\mathbf r\, \mathbf J(\mathbf r,t)\cdot\nabla\mu(\mathbf r,t).
\label{eq:mudrho}
\end{align}
Under no-flux boundary conditions, the divergence term reduces to a surface contribution that vanishes, yielding
\begin{equation}
\int d\mathbf{r}\mu\partial_t\rho
=
\int d\mathbf r\, \mathbf J\cdot\nabla\mu.
\label{eq:mudrho2}
\end{equation}

Using the DDFT constitutive relation for the current, $\mathbf J = -D\rho\nabla(\beta\mu)$, the first contribution can be rewritten as
\begin{equation}
\int d\mathbf r\, \mu\,\partial_t\rho=-k_BT \int d\mathbf r\,
\frac{|\mathbf J(\mathbf r,t)|^2}{D\,\rho(\mathbf r,t)}=-\dot{S}(t)T,
\label{eq:mudrho3}
\end{equation}
which has been written in terms of the entropy production rate, defined as
\begin{equation}
\frac{\dot{S}(t)}{k_\mathrm{B}}= \int d\mathbf r\,
\frac{|\mathbf J(\mathbf r,t)|^2}
{D\,\rho(\mathbf r,t)} = A\int_R^{L(t)-R} dz\,
\frac{J(z,t)^2}
{D\,\rho(z,t)},
\label{eq:entropyproduction}
\end{equation}

$\dot{S}(t)$ represents the instantaneous irreversible entropy production of the combined system (colloidal fluid and thermal bath) within the DDFT approximation~\cite{Schmidt2022}, generated by the dissipative diffusive currents driven by the mobile piston, and is thus strictly non-negative~\cite{Seifert2005EntropyProduction}.

We now turn to the second contribution appearing in Eq.~\eqref{eq:dFdt2}, which arises from the explicit time dependence of the external potential through the piston position. This term reads
\begin{equation}
\int d\mathbf r\, \rho(\mathbf r,t)\,
\partial_t V_{\mathrm{ext}}(\mathbf r,t)
=
A \int_R^{L(t)-R} dz\, \rho(z,t)\,
\frac{\partial V_{\mathrm{ext}}(z,t)}{\partial t},
\label{eq:rhoVextdot}
\end{equation}

We consider a wall confinement in which only the right wall (the piston)
contributes to the explicit time dependence of the external potential via the
piston position $L(t)$. Accordingly,
\begin{equation}
\partial_t V_{\mathrm{ext}}(z,t)=\dot L(t)\,\frac{\partial V_{\mathrm R}}{\partial L}=
-\dot L(t)\,\frac{\partial V_{\mathrm R}}{\partial z},
\label{eq:Vextdot}
\end{equation}
where the second equality follows from the dependence
$V_{\mathrm R}=V_{\mathrm R}(L(t)-z)$.

Substituting Eq.~\eqref{eq:Vextdot} into Eq.~\eqref{eq:rhoVextdot} and using the
definition of the instantaneous pressure exerted by the fluid on the piston,
Eq.~\eqref{eq:PR}, we obtain
\begin{equation}
\int d\mathbf r\, \rho\,\partial_t V_{\mathrm{ext}}
=
A \dot L(t)\int dz\, \rho
\frac{\partial V_{\mathrm R}}{\partial z}
=
- A P_{\mathrm R}(t)\,\dot L(t).
\label{eq:power_term}
\end{equation}

Here $\dot L(t)$ denotes the instantaneous piston velocity. Eq.~\eqref{eq:power_term} therefore represents the instantaneous mechanical power injected into the colloidal fluid by the time-dependent external potential,
\begin{equation}
\dot W(t) = - A P_{\mathrm R}(t)\,\dot L(t).
\label{eq:mechpower_def}
\end{equation}

Using the overdamped piston dynamics,
Eq.~\eqref{eq:piston_overdamped}, the mechanical power can be expressed in terms
of the piston mobility $K$ and the imposed external pressure,
\begin{equation}
\dot W(t)
=
- A K\,P_{\mathrm R}(t)\,
\bigl(P_{\mathrm R}(t) - P_{\mathrm{ext}}\bigr).
\label{eq:mechpower_K}
\end{equation}

Combining Eqs.~\eqref{eq:mudrho3} and \eqref{eq:rhoVextdot} into
Eq.~\eqref{eq:dFdt2}, we finally arrive at
\begin{equation}
\frac{d\,\beta\mathcal F[\rho;t]}{dt}
=
-\frac{\dot S(t)}{k_B}
+
\beta\,\dot W(t),
\label{eq:dFdt_general}
\end{equation}
which explicitly separates the strictly dissipative contribution (first term)
from the rate at which the external driving modifies the free energy through the motion of the piston (second term). Equation~\eqref{eq:dFdt_general} shows that in the absence of external driving
($\partial_t V_{\mathrm{ext}}=0$) the total free energy is a Lyapunov functional and decreases monotonically. The piston mobility $K$ controls the rate of this energy exchange via $\dot L(t)$ and therefore determines how strongly the system is driven out of equilibrium during the transient compression.

Integrating Eq.~\eqref{eq:dFdt_general} from the initial equilibrium state at
$t=0$ to the final equilibrium state at $t=t_\infty$ yields the difference in (Helmholtz) equilibrium free energy~\cite{jarzynski1997nonequilibrium, crooks1999entropy}, 
\begin{equation}
\beta\Delta\mathcal F_{\mathrm{eq}}=-\frac{\Delta S}{k_B}+\beta\Delta W,
\label{eq:integrated_balance}
\end{equation}
where
\begin{equation}
\Delta S=\int_0^{t_\infty} dt\,\dot{S}(t),
\label{eq:DS}
\end{equation}
is the total entropy produced during the non-equilibrium compression process, and
\begin{align}
\Delta W=&\int_0^{t_\infty} dt\,\dot W(t)= -A\int_0^{t_\infty}dt\,P_\mathrm{R}(t)\dot{L}(t)\nonumber \\
=&A\int_{L_0}^{L_\infty} P_\mathrm{R}(L)dL.
\label{eq:PRL}
\end{align}
is the total mechanical work transferred by the piston on the colloidal fluid.

\subsection{Physical interpretation of $\Delta S$}

It is instructive to clarify the physical meaning of the entropy production term, $\Delta S$, introduced in Eq.~\eqref{eq:entropyproduction}. In non--equilibrium thermodynamics, the total entropy change of the universe can be written as the sum of the entropy variation of the system and that of the surrounding thermal bath~\cite{Seifert_2012},
\begin{equation}
\Delta S = \Delta S_{\mathrm{sys}} + \Delta S_{\mathrm{bath}}.
\label{eq:SsysSbath}
\end{equation}
In the present problem, the ``universe'' consists of the confined colloidal fluid system and the thermal reservoir that maintains the temperature constant. Within DDFT, $\Delta S$ corresponds to the irreversible entropy production and therefore satisfies $\Delta S \ge 0$, in accordance with the second law of thermodynamics.

If no particle currents are present, the system remains arbitrarily close to equilibrium and $\Delta S=0$. This occurs in the quasi-static limit of very slow compression ($K\to 0$), where the density profile continuously adapts to the instantaneous piston position and no dissipative fluxes develop. In contrast, for rapid compression ($K\gg 1$), significant particle currents arise and $\Delta S$ measures the irreversible entropy generated by the diffusive relaxation of the fluid.

The above identity can be used to determine the entropy change of the thermal bath. On the one hand, from the energetic balance given by Eq.~(\ref{eq:integrated_balance}) one has $\Delta \mathcal{F}_{\mathrm{eq}} = -T \Delta S + \Delta W$. On the other hand, in equilibrium the free energy for hard spheres is decomposed into an intrinsic (purely entropic) contribution and an external potential energy term,
$\mathcal{F}_{\mathrm{eq}} = -T S_{\mathrm{sys}} + U_{\mathrm{ext}}$, with $-T S_{\mathrm{sys}} = \mathcal{F}_{\mathrm{id}} + \mathcal{F}_{\mathrm{ex}}$ and 
$U_{\mathrm{ext}} = \int d\mathbf{r}\,\rho(\mathbf{r}) V_{\mathrm{ext}}(\mathbf{r})$.
Therefore,
\begin{equation}
\Delta \mathcal{F}_{\mathrm{eq}} = -T \Delta S_{\mathrm{sys}} + \Delta U_{\mathrm{ext}}.
\label{eq:Feq2}
\end{equation}

Physically, compression reduces the configurational entropy of the colloidal system ($\Delta S_{\mathrm{sys}}<0$), reflecting the decrease in the number of accessible microstates (through excluded volume), while $\Delta U_{\mathrm{ext}}>0$ increases due to particle accumulation near the confining walls. Equating both expressions for $\Delta \mathcal{F}_{\mathrm{eq}}$ and using Eq.~\eqref{eq:SsysSbath} one obtains
\begin{equation}
T \Delta S_{\mathrm{bath}} = \Delta W - \Delta U_{\mathrm{ext}}.
\label{eq:Sbath}
\end{equation}

In the quasi-static limit ($K\to 0$), $\Delta S=0$ and the entropy decrease of the system is exactly compensated by the entropy increase of the bath, which takes its minimum possible value consistent with reversibility. For large $K$, the entropy production becomes substantial and the bath entropy increases accordingly, as it absorbs the heat generated by irreversible diffusive currents in the fluid.

\section{Numerical procedure and calculation of observables}
\label{sec:numerical}

The DDFT-piston coupled dynamics implies the solution to the DDFT equations and the piston equation (\ref{eq:piston_overdamped}) self-consistently in the quasi-1D planar setup, that is, in coordinates $z$ and $t$. A convenient choice for the interaction of the particles with the walls ($V_{\rm ext}=V_\mathrm{R}+V_\mathrm{L}$) is provided by a short-ranged, smoothly varying repulsive potential of the form
\begin{align}
\label{eq:VL}
V_\mathrm{L}(z)=&V_0\left(1-\tanh(\lambda(z-R))\right) \\
V_\mathrm{R}(z,L(t))=&V_0\left(1-\tanh(\lambda(L(t)-z-R))\right).
\label{eq:VR}
\end{align}
The
parameters $V_0$ and $\lambda$ control the strength and spatial range of the
repulsion, respectively.  Throughout this work we choose $\beta V_0 = 50$ and $\lambda R = 5$.


The local fluid velocity can be obtained as $v(z,t)=J(z,t)/\rho(z,t)$. 
The material (particle-weighted) average velocity of the fluid is then given by
\begin{equation}
\langle v(t) \rangle =\frac{A}{N}\int_0^{L(t)} dz\, J(z,t)
\label{eq:v_material}
\end{equation}

In order to present the results in a compact and universal form, all physical
quantities are expressed in terms of dimensionless variables. As fundamental
scales we choose the particle radius $R$ as the unit of length, the Brownian
time $\tau_B = R^2/D$ as the unit of time, and thermal energy $k_\mathrm{B}T$ as the unit of energy. Spatial coordinates and time are non-dimensionalized as $z^* = z/R$ and $t^* = t/\tau_\mathrm{B}$, respectively. The number density and particle current are expressed in reduced form as $\rho^*(z,t) = \rho(z,t)\,R^3$ and 
$J^*(z,t) = R^2\tau_\mathrm{B}J(z,t)$ The mean fluid velocity is written as $\langle v^*(t) \rangle = (\tau_\mathrm{B}/R)\,\langle v(t) \rangle$. Pressures are non-dimensionalized using the natural thermal pressure scale, 
$P^* = \beta R^3 P$. The instantaneous mechanical power injected by the piston per unit of area, $\dot{W}(t)/A$, is rendered dimensionless as $\dot{W}^*(t)=(\beta\tau_\mathrm{B}R^2)(\dot{W}(t)/A)$. For the total mechanical work per unit of area $\Delta W^*=(\beta R^2)(\Delta W/A)$. The reduced entropy production rate and total produced entropy per unit of area are $\dot{S}^*/k_\mathrm{B}=(\tau_\mathrm{B}\beta R^2)(T\dot{S}/A)$, and $\Delta S^*/k_\mathrm{B}=\beta R^2(T\Delta S/A)$. Analogously, $\Delta \mathcal{F}^*_\mathrm{eq}=(\beta R^2)(\Delta\mathcal{F}/A)$ and $\Delta U^*_\mathrm{ext}=(\beta R^2)(\Delta U_\mathrm{ext}/A)$. Finally, the reduced piston mobility is $K^*=(\tau_\mathrm{B}/(\beta R^4))K$.

To solve our set of equations numerically, space is discretized on a uniform grid with spacing $\Delta z^* = 0.01$. Time is discretized using a constant time step $\Delta t^* = 10^{-6}$. The DDFT equation for the density field $\rho^*(z,t)$ is integrated forward in time using an explicit finite-difference scheme, while the piston position $L^*(t)$ is updated at each time step according to the overdamped equation of motion. No-flux boundary conditions, $J^*(R,t)=J^*(L_0-R,t)=0$, are imposed throughout the whole process.

The system is prepared in an equilibrium state under an initial reduced piston pressure $P_0^* = 0.128$ and with a uniform mean reduced particle density $\rho_0^* = 0.05$. For an initial reduced wall separation $L_0^* = 10$, and accounting for the excluded regions adjacent to the confining walls, this corresponds to an effective accessible length $(L_0-2R)$ and to a number of particles per unit transverse area $N/A = \rho_0 (L_0-2R) = 0.4R^{-2}$. The quantity $N/A$ is conserved throughout the dynamics. The corresponding equilibrium density profile $\rho^*(z,0)$ is obtained by solving the static Euler--Lagrange equation associated with the DFT functional.

At $t=0$, the pressure acting on the piston is abruptly switched to a larger value, $P_{\mathrm{ext}}^* = 0.5$, thereby inducing a compression of the fluid. The system then evolves dynamically toward a new equilibrium state consistent with the imposed pressure, characterized by $P_\infty^* = P_{\mathrm{ext}}^* = 0.5$ and $L_\infty^*= 6.158$. The system is considered to have reached equilibrium when the change in the reduced density profile between two successive time steps satisfies $\int dz^*\bigl[\rho^*(z,t_{n+1}) - \rho^*(z,t_n)\bigr]^2< 10^{-24}$. 

Calculations are performed for a wide range of mobilities, $0.01 \le K^* \le 10000$, covering regimes in which the piston motion is much slower than, comparable to, or much faster than the intrinsic Brownian relaxation of the colloidal fluid. For each value of $K^*$, we record the time evolution of $\rho^*(z,t)$, $J^*(z,t)$, $P_\mathrm{L}^*(t)$, $P_\mathrm{R}^*(t)$, $L^*(t)$, $\langle v^*(t) \rangle$, $\dot {W}^*(t)$ and $\dot{S}^*(t)$. In addition, we compute the time-integrated quantities $\Delta W^*$ and $\Delta S^*$. 


\section{Results and discussion}
\label{sec:results}

\begin{figure}[ht!]
	\centering
	\includegraphics[width=1\linewidth]{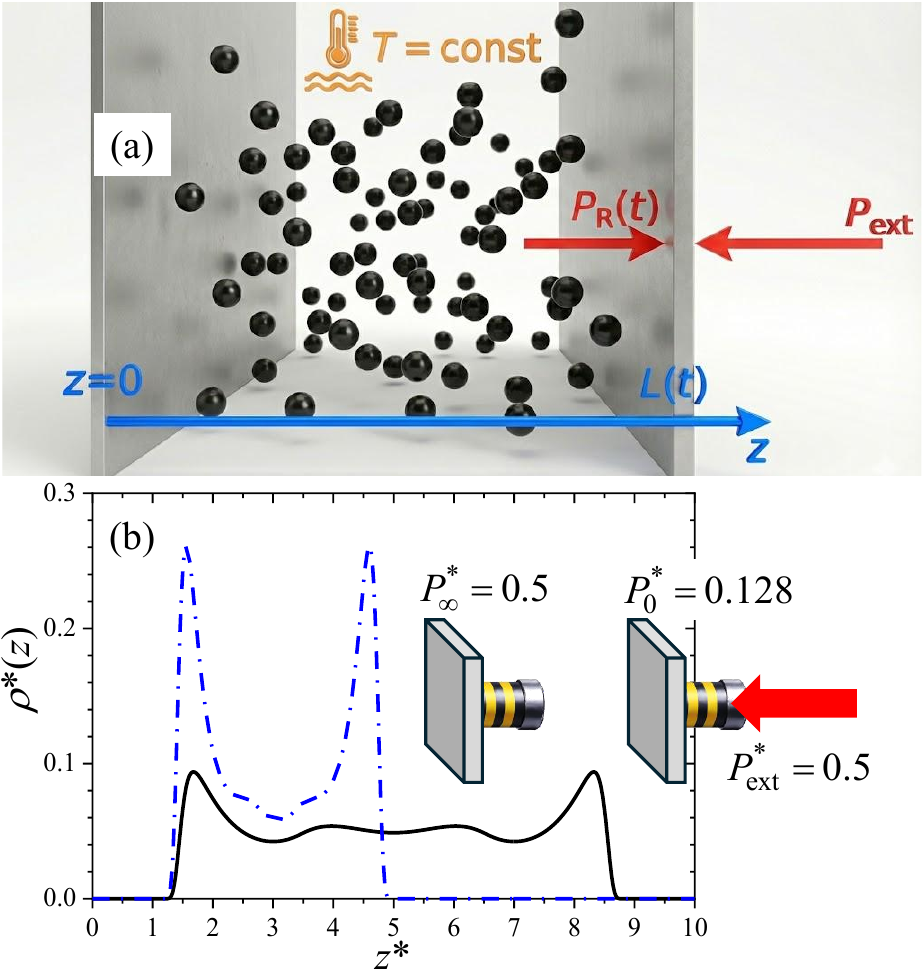}
	\caption{(a) Schematic illustration of a hard-sphere fluid confined between two parallel walls. The right boundary is an overdamped mobile piston subjected to an externally imposed pressure $P_{\mathrm{ext}}$. (b) Equilibrium density profiles of the hard-sphere fluid in the initial and final states (solid black and dashed blue lines, respectively), corresponding to an initial piston pressure $P_0^* = 0.128$ with wall separation $L^*_0 = 10$, and to the final compressed equilibrium state under $P^*_{\mathrm{ext}} = P_\infty = 0.5$ with final piston position $L^*_\infty = 6.158$, respectively. In both cases the reduced mean particle density is $\rho^* = 0.05$.}
	\label{fig:Fig1}
\end{figure}

Figure~\ref{fig:Fig1}(a) schematically illustrates the one--dimensional hard--sphere fluid confined between a fixed left wall and a mobile piston on the right. The piston dynamics is governed by the balance between the externally applied reduced pressure $P_\mathrm{ext}^*$ and the instantaneous pressure exerted by the confined fluid. The system is driven out of equilibrium by a sudden change of the reduced external pressure from its initial value $P_0^*$ to a larger final value $P_\mathrm{ext}^*$, which triggers a time--dependent motion of the piston.

The black solid line in Fig.~\ref{fig:Fig1}(b) shows the equilibrium density profile corresponding to the initial state at $P_0^* = 0.128$ for the fluid within length $L_0^* = 10$.  The density profile exhibits the characteristic oscillatory modulations of a confined hard--sphere system, with pronounced layering close to both confining boundaries. In particular, four well--defined coordination layers are observed within the accessible region $[R,\,L_0-R]$.

After evolution and equilibration under the larger external pressure $P_\mathrm{ext}^* = 0.5$, the piston reaches a smaller stationary separation $L_\infty^* = 6.158$, resulting in a denser packing of the fluid and a clear reorganization of the layering structure, shown as a dashed blue line in Fig.~\ref{fig:Fig1}(b). Compared to the initial state, the density peaks are shifted and amplified, forming two dense regions near both walls. The difference in the equilibrium Helmholtz free energy per unit area between the final and initial states is  $\Delta \mathcal{F}_\mathrm{eq}^* = 0.980$. The corresponding change of entropy and external potential energy are given by $\Delta S_\mathrm{sys}^*/k_\mathrm{B}=-0.904$ and $\Delta U_\mathrm{ext}^*=0.076$, respectively.


\subsection{Non-equilibrium density profiles during compression}

To characterize the non-equilibrium response, we introduce the timescale $\tau_p$ associated with piston motion over a distance of order the particle radius $R$. From the piston equation of motion, $R/\tau_p = K \left(P_\mathrm{ext}^* - P_0^*\right)$, so that the ratio between Brownian and piston timescales becomes
$\gamma \equiv R^2/(2D\tau_\mathrm{p}) =\tau_\mathrm{B}/(2\tau_p) = K^*\left(P_\mathrm{ext}^* - P_0^*\right)/2 \simeq 0.2K^*$, 
where the last estimate corresponds to the pressure jump used here. Thus, for $K^* \gtrsim 5$ the piston moves faster than the colloidal fluid relaxes diffusively, whereas for $K^* \lesssim 5$ diffusion remains faster than piston motion.

\begin{figure}[ht!]
	\centering
	\includegraphics[width=1\linewidth]{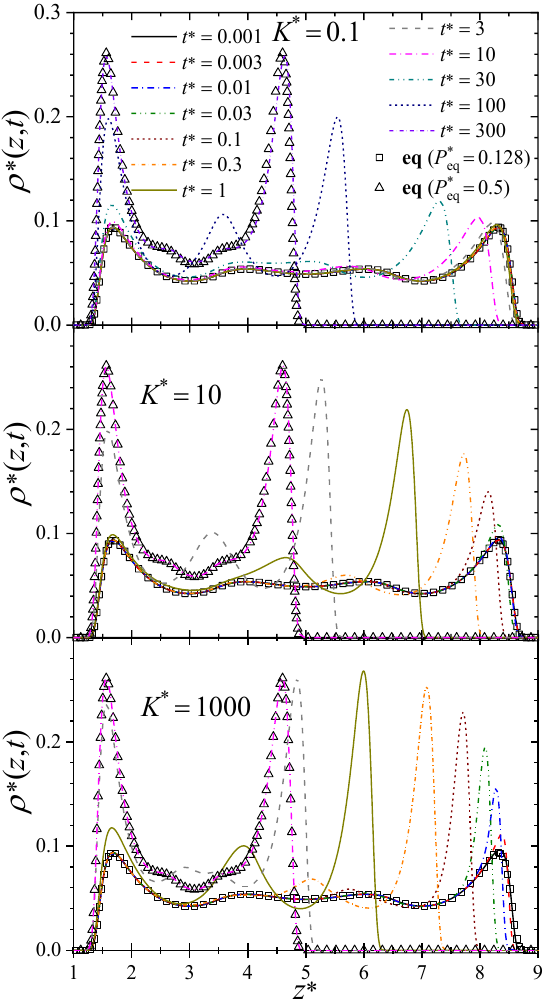}
	\caption{(a) Time evolution of the non-equilibrium density profiles during the compression process. From top to bottom, the panels correspond to $K^* = 0.1$, $10$, and $1000$. In the top panel, square and triangle symbols indicate the limiting equilibrium density profiles obtained at reduced pressures $P_\mathrm{R}^* = 0.128$ and $P_\mathrm{R}^* = 0.5$, respectively.
}
	\label{fig:Fig2}
\end{figure}

Fig.~\ref{fig:Fig2}(a) shows the non-equilibrium density profiles $\rho^*(z,t)$ during the compression process at different reduced times ranging from $t^* = 10^{-3}$ to $t^* = 300$, for three representative piston mobilities, $K^* = 0.1$, $10$, and $1000$. For reference, the equilibrium density profiles corresponding to the initial and final states are also shown as square and triangular symbols, respectively.

For the slow piston, $K^* = 0.1$, the density profiles evolve smoothly in time, displaying a gradual amplification of the density peaks as the piston slowly moves leftwards and induces a gentle compression of the fluid. The final equilibrium state is reached only at long times, $t^* \approx 300$. In this regime, the compression process is effectively quasi-static. This is in particularly reflected in the nearly symmetric heights of the density peaks near the left and right confining walls, indicating that the colloidal particles have sufficient time to redistribute throughout the entire system. 

For the intermediate mobility $K^*=10$, the compression process becomes significantly faster and the system reaches the final equilibrium state already at $t^* \approx 10$. In contrast to the quasi-static regime, a pronounced asymmetry develops between the density peaks near the left and right walls, signaling that the system is driven significantly away from equilibrium. In particular, a noticeable accumulation of particles builds up in the vicinity of the piston due to the advective dragging induced by its motion.

In the limit of very large piston mobility, $K^* = 1000$, the piston responds almost instantaneously to the applied pressure jump and rapidly reduces the pressure imbalance. This leads to a sharp increase of the density peak near the piston at very short times, which can even exceed the height of the corresponding peak in the final equilibrium compressed state. A strong asymmetry between the density profiles near the left and right walls is observed at early times. At later stages, however, the colloidal fluid undergoes a diffusive readjustment, progressively reducing the excess particle concentration near the piston until the density peaks at both walls become comparable. Remarkably, despite the piston being two orders of magnitude faster than in the $K^* = 10$ case, the final equilibrium state is reached on a similar timescale, $t^* \approx 10$. This observation highlights that, beyond a certain piston mobility threshold, the overall relaxation process is no longer limited by the piston dynamics but rather by the intrinsic diffusive timescale of the colloidal fluid.

The non-equilibrium current profiles $J^*(z,t)$ are most pronounced in the vicinity of the mobile piston and, as $K$ increases, tend to saturate, ultimately being bounded by the intrinsic diffusive timescale of the colloids (see Appendix A for further discussion).

\begin{figure*}[ht!]
	\centering
	\includegraphics[width=1\linewidth]{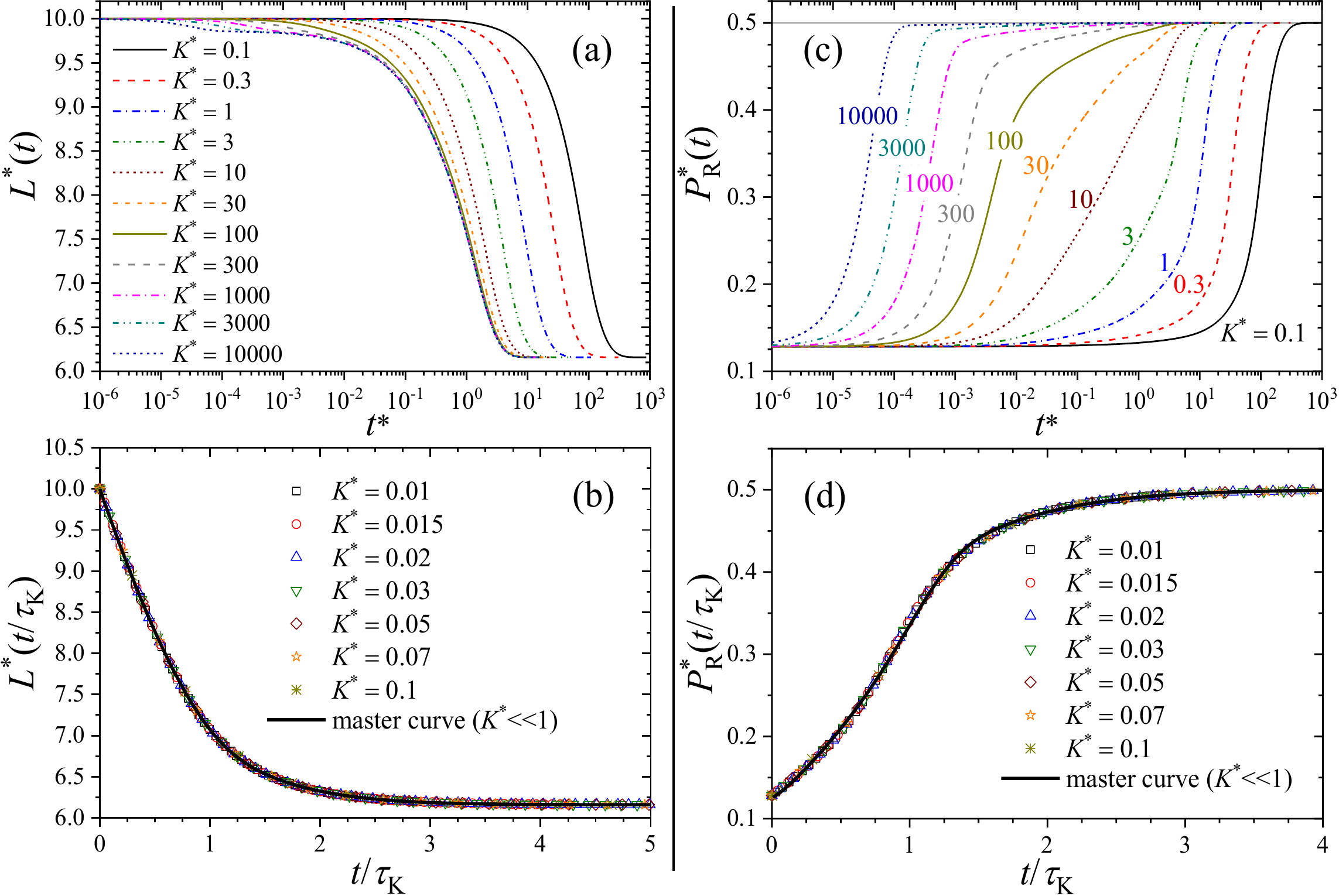}
	\caption{(a) Time evolution of the distance between walls during the compression process, $L^*(t)$, for values of the mobility parameter $K^*$ ranging from $0.1$ to $10000$. (b) Collapse of $L^*(t)$ onto a common master curve when plotted as a function of the normalized time $t/\tau_{\mathrm{K}}$ in the regime $K^* \ll 1$. The solid black line represents the theoretical master curve given by Eq.~\eqref{eq:lamejorfuncion}. (c) Time evolution of the pressure exerted by the hard-sphere fluid on the mobile hard wall, $P_\mathrm{R}^*(t)$, for values of $K^*$ between $0.1$ and $10000$. (d) Scaling collapse of $P_\mathrm{R}^*(t)$ as a function of $t/\tau_{\mathrm{K}}$ in the regime $K^* \ll 1$, yielding a universal curve described by Eq.~\eqref{eq:lasegundamejorfuncion}.
    }
	\label{fig:Fig3}
\end{figure*}

\subsection{Macroscopic dynamics: piston position, pressure, and mean velocity}


Fig.~\ref{fig:Fig3}(a) shows the time evolution of the reduced piston position $L^*(t)$ for piston mobilities ranging from $K^* = 0.1$ up to $K^* = 10000$ plotted over more than six decades in time. 


In the low--mobility regime, the curves corresponding to different values of $K^*$ display very similar shapes but are shifted along the time axis, suggesting the existence of a universal scaling behavior. This observation is confirmed in Fig.~\ref{fig:Fig3}(b), where the piston position is plotted as a function of the rescaled time $t/\tau_K$ on a linear scale, with the characteristic piston timescale defined as
\begin{equation}
\tau_K = \frac{L_0-L_\infty}{K(P_\mathrm{ext}-P_0)},
\label{eq:tauK}
\end{equation}
so that $\tau_K^*\equiv \tau_K/\tau_\mathrm{B}=10.337/K^*$. Indeed, all curves corresponding to small piston mobilities, $0.01 \le K^* \le 0.1$, collapse onto a single master curve. This universal behavior can be fitted by the functional form
\begin{align}
L(t)=&L_0-(L_0-L_\infty)\Big\{ C\big[1-\exp\big(-(t/\tau_K)^a\big)\big]^{1/a}  \nonumber \\
+&(1-C)\big[1-\exp\big(-(t/\tau_K)^b\big)\big]^{1/b} \Big\}.
\label{eq:lamejorfuncion}
\end{align}

This expression satisfies the correct asymptotic limits, namely $L(t)\to L_\infty$ for $t\to\infty$, while the parameter $a$ controls the curvature of the relaxation toward the final state. At very short times, the expression reduces to a linear decay, $L(t)\simeq L_0-(L_0-L_\infty)(t/\tau_K)$, yielding an initial piston velocity $(\mathrm{d}L/\mathrm{d}t)_0=-(L_0-L_\infty)/\tau_K$. Using the piston equation of motion, $(\mathrm{d}L/\mathrm{d}t)_0=K(P_0^*-P_\mathrm{ext}^*)$, one consistently recovers the definition of $\tau_K$. The solid black line in Fig.~3(b) corresponds to this fit, which is excellent for $a = 1.33$, $b = 3.91$ and $C=0.703$. The description of the relaxation dynamics using a double stretched exponential form suggests the presence of a broad and possibly heterogeneous spectrum of relaxation times, arising from the interplay between diffusive transport and boundary-driven compression.

As $K^*$ increases, the scaling behavior progressively breaks down and a qualitatively different dynamical regime emerges. For sufficiently large values of $K^*$, the piston dynamics exhibits a particularly interesting behavior. In this high--mobility regime, the piston motion proceeds through two distinct temporal stages. During the first stage, the piston rapidly moves from the initial position $L_0^* = 10$ to an intermediate position $L_1^* = 9.857$, with an initial velocity that increases with the piston mobility. This fast relaxation stage can be accurately described by a similar but simpler functional form as in Eq.~\eqref{eq:lamejorfuncion}, namely
\begin{equation}
L(t)=L_0-(L_0-L_1)\big[1-\exp\big(-(t/\tau_K')^{a'}\big)\big]^{1/a'},
\end{equation}
with a curvature exponent $a' = 1.364$ and
\begin{equation}
    \tau_K'=\frac{L_0-L_1}{K(P_\mathrm{ext}-P_0)},
    \label{eq:tauKp}
\end{equation}
leading to $\tau_K'^*=0.384/K^*$.

Once this intermediate state is reached, the system enters a second relaxation stage, visible in Fig.~\ref{fig:Fig3}(a) as a small shoulder in the piston trajectory. During this stage, the piston evolves slowly toward the final equilibrium position $L_\infty^*$, following the same saturation curve observed for all large $K^*$. This saturation reflects again the dynamical limiting behavior imposed by the intrinsic diffusion of the colloids.

Fig.~\ref{fig:Fig3}(c) shows the reduced right--wall pressure exerted by the piston $P_\mathrm{R}^*(t)$ as a function of time for piston mobilities in the range $0.1 \le K^* \le 10000$. As expected, the pressure increases toward its new equilibrium value more rapidly as the piston mobility is increased. When considering all curves together, it is apparent that for small values of $K^*$ the temporal evolution of the pressure follows a very similar functional form, suggesting the existence of a scaling behavior.

This scaling is clearly demonstrated in Fig.~\ref{fig:Fig3}(d), where the pressure curves corresponding to $0.01 \le K^* \le 0.1$ collapse onto a single master curve when plotted as a function of the normalized time $x=t/\tau_K$, with $\tau_K$ given by Eq.~\ref{eq:tauK}. Using the piston equation of motion, Eq.~\eqref{eq:piston_overdamped}, we find that $P_\mathrm{R}(t)=P_\mathrm{ext}+(1/K)\,\mathrm{d}L/\mathrm{d}t$. Substituting the fitted expression for $L(t)$ given by Eq.~\eqref{eq:lamejorfuncion} yields the explicit form
\begin{align}
&\frac{P_\mathrm{R}(t)-P_\mathrm{ext}}{P_0-P_\mathrm{ext}}=  Cx^{a-1}\exp(-x^a)\big[1-\exp(-x^a)\big]^{\frac{1}{a}-1} \nonumber \\
&\ \ \ \ \ +(1-C)x^{b-1}\exp(-x^b)\big[1-\exp(-x^b)\big]^{\frac{1}{b}-1},
\label{eq:lasegundamejorfuncion}
\end{align}
which accurately reproduces the collapsed data, as shown by the solid black line in Fig.~\ref{fig:Fig3}(d).

As the piston mobility increases, this scaling behavior is progressively lost. Approaching the high-mobility saturation regime, the piston rapidly adjusts its position such that the pressure exerted by the fluid nearly matches the externally imposed pressure. This is illustrated by the curve corresponding to $K^* = 10000$, for which $P_\mathrm{R}^*(t)$ rises abruptly and reaches the external pressure on a very short timescale, $t^* \approx 10^{-4}$. This rapid pressure buildup is enabled by the strong advective dragging of the colloidal fluid, which leads to a transient accumulation of particles near the piston, as previously evidenced by the density profiles in Fig.~\ref{fig:Fig2}(a). Beyond this initial stage, the subsequent evolution of the pressure is governed by the diffusive redistribution of the colloidal fluid. In this regime, $P_\mathrm{R}(t)$ follows a universal saturation curve given by
\begin{equation}
\frac{P_\mathrm{R}(t)-P_\mathrm{ext}}{P_0-P_\mathrm{ext}}=x^{a'-1}\exp(-x^{a'})\big[1-\exp(-x^{a'})\big]^{\frac{1}{a'}-1},
\label{eq:laterceramejorfuncion}
\end{equation}
when written in terms of the rescaled time $x'=t/\tau_K'$, with $\tau_K'$ given by Eq.~\eqref{eq:tauKp} and characterized by the exponent $a' = 1.364$

While the piston timescale is set by the mobility $K$, the confined fluid relaxes through a spectrum of diffusive modes \cite{marconi1999dynamic,archer2004}. These range from rapid local density rearrangements near the walls to slower, global particle redistributions governed by the confinement geometry \cite{RexLowen2008}. This hierarchy dictates the observed dynamics: in the low-mobility regime, the piston evolves slowly enough ($\tau_K \gg \tau_{\text{fluid}}$) for the fluid to remain in local equilibrium, leading to quasi-static scaling \cite{SchmidtBrader2003}. Conversely, at high mobility, the piston rapidly compresses the first solvation layer---manifesting as a transient shoulder in the dynamics---before being limited by the slower diffusive restructuring of the inner layers. This separation of timescales explains the emergence of universal saturation curves, where long-time relaxation becomes independent of piston mechanics and is governed solely by the intrinsic transport properties of the confined fluid \cite{almenar2011dynamics}.

\begin{figure}[ht!]
	\centering
	\includegraphics[width=1\linewidth]{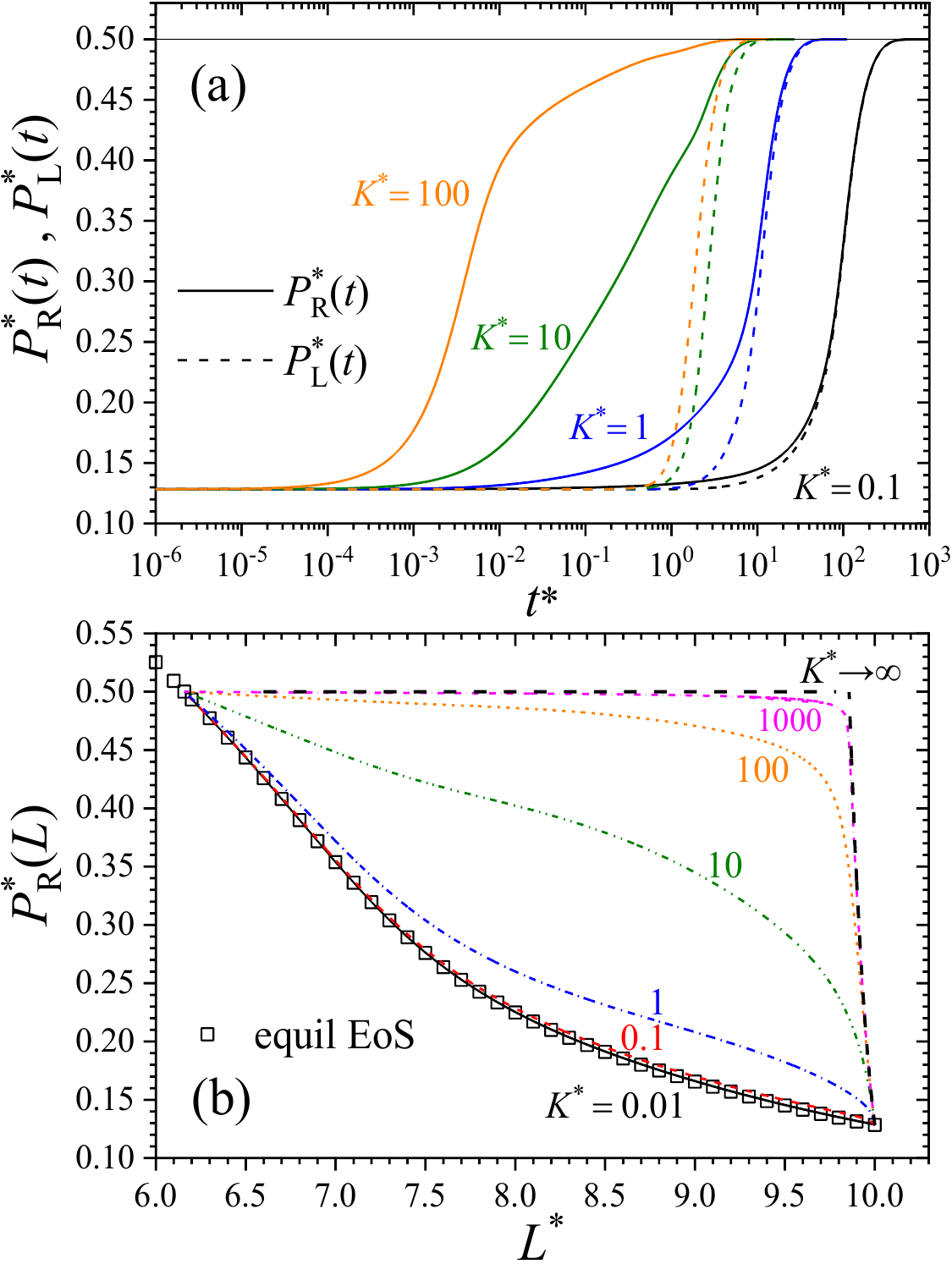}
	\caption{(a) Comparison between the time evolution of the pressure on the hard and left walls. (b) Pressure on the hard wall as a function of the distance between walls, $P_\mathrm{R}^*(L)$ for values of $K^*$ from $0.01$ to $1000$ ($P_0^*=0.128$ and $P_\mathrm{ext}^*=0.5$). Square symbols represent the equilibrium equation of state (EoS) of the confined colloids obtained from equilibrium DFT, $P_\mathrm{eq}^*(L^*)$. Black dashed line represents the limiting behavior obtained for pistons with very high piston mobility ($K^*\rightarrow \infty$).}
	\label{fig:Fig4}
\end{figure}


Fig.~\ref{fig:Fig4}(a) now compares the temporal evolution of the asymmetric pressures exerted by the fluid on the right wall (the mobile piston) versus the fixed left wall, shown as solid and dashed lines, respectively, for piston mobilities in the range $0.1 \le K^* \le 100$. For $K^* = 0.1$ the two pressures almost coincide, as expected near equilibrium, except for very short initial transients at the onset of the compression process. 

As the piston mobility is increased, a significant time lag develops between the pressures at the two walls. In particular, the pressure at the left wall, $P_\mathrm{L}(t)$, responds increasingly slowly to the piston motion, reflecting the finite time required for the compression to propagate through the system. For instance, at $K^* = 100$, the delay between $P_\mathrm{R}(t)$ and $P_\mathrm{L}(t)$ spans almost two decades in time, providing a clear signature of strongly non-equilibrium dynamics controlled by diffusive transport across the slit.

Fig.~\ref{fig:Fig4}{(b) presents the pressure exerted by the fluid on the piston, $P_\mathrm{R}(L)$, as a function of the instantaneous piston position $L$, for piston mobilities ranging from $K^* = 0.01$ to $K^* = 1000$. This representation is particularly informative, since the area enclosed by the curve directly yields the mechanical work per unit area injected by the piston into the fluid, $\Delta W/A$, according to Eq.~\eqref{eq:PRL}. For reference, square symbols indicate the equilibrium pressure $P_\mathrm{eq}(L)$ for each piston position, corresponding to the equilibrium equation of state (EoS) of the confined colloidal fluid.

For the smallest mobility, $K^* = 0.01$, the curve $P_\mathrm{R}(L)$ follows the EoS almost exactly, as expected for a quasi--static process. In this limit, all the injected work is reversibly converted into an increase of the Helmholtz free energy, yielding $\lim_{K\rightarrow 0}\Delta W^* = \Delta F_\mathrm{eq}^* = 0.980$. This value therefore represents a lower bound for the work performed on the fluid.

As $K^*$ increases, the curves progressively deviate from the EoS, indicating the onset of irreversible dynamics and enhanced work injection. Importantly, however, the injected work does not grow without bound as the piston mobility increases. Instead, the curves converge asymptotically toward a limiting behavior, shown by the dashed black line, corresponding to the idealized limit $K^* \to \infty$. Using the analytical fits previously obtained for $L(t)$ and $P_\mathrm{R}(t)$ in the high--mobility regime, it is possible to eliminate the explicit time dependence and express $P_\mathrm{R}$ directly as a function of $L$, namely
\begin{equation}
\lim_{K \to \infty} P_\mathrm{R}(L) = P_\mathrm{ext},
\qquad L_\infty \le L \le L_1 .
\label{eq:PRKinf_1}
\end{equation}
and
\begin{equation}
\begin{alignedat}{1}
\lim_{K \to \infty} P_\mathrm{R}(L)
= P_\mathrm{ext}
- (P_\mathrm{ext} - P_0)
(1-y^{a'})\,y^{1-a'}
\\
\times\big[-\ln(1-y^{a'})\big]^{1-1/a'},
\qquad L_1 < L \le L_0 .
\end{alignedat}
\label{eq:PRKinf_2}
\end{equation}
where $y(L)=(L_0-L)/(L_0-L_1)$.

Evaluating the enclosed area yields a total injected work per unit of area $\Delta W^* = 1.89$, which constitutes an upper bound for the work performed by the piston on the fluid.

\begin{figure}[ht!]
	\centering
	\includegraphics[width=1\linewidth]{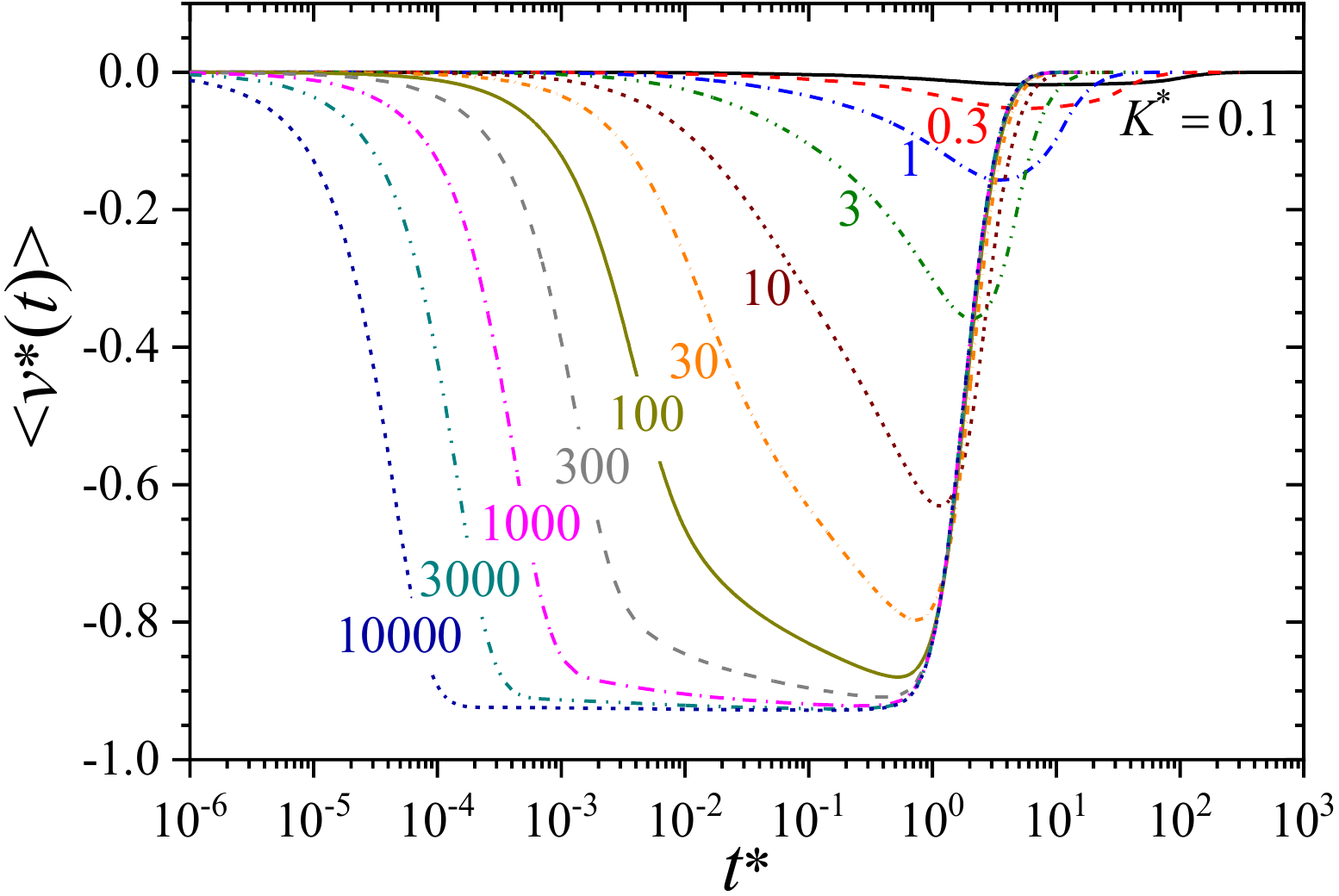}
	\caption{Time evolution of the mean particle velocity for values of $K^*$ ranging from $0.1$ to $10000$. Negative values indicate that particles move, on average, toward the left as a result of the action of the compressing piston.
}
\label{fig:Fig5}
\end{figure}

Fig.~\ref{fig:Fig5} shows the time evolution of the mean reduced velocity of the colloidal particles, $\langle v^*(t) \rangle$, during the compression process, for piston mobilities ranging from $K^* = 0.1$ to $K^* = 10000$. Upon activation of the compression protocol, the magnitude of the velocity increases from the initial zero, reaching a maximum, and subsequently decays back to zero as the system approaches the final equilibrium state. 

For slow piston mobility, $K^* = 0.1$, the mean particle velocity remains very small throughout the process and deviates appreciably from zero only at times of the order of $t^* \sim 1$. As the piston mobility increases, the velocity profiles become progressively more pronounced and the minimum in $\langle v^*(t) \rangle$ is reached at increasingly earlier times, reflecting the faster mechanical driving imposed by the piston. For $K^* \gtrsim 30$, the velocity dynamics exhibits a qualitatively new feature: an intermediate temporal regime emerges in which the velocity decreases approximately logarithmically with time. Interestingly, this regime becomes broader and flattens as $K^*$ increases, i.e., an extended time window develops in which the colloidal fluid moves with an approximately constant upper limiting velocity, $\langle v^*(t) \rangle \simeq -0.924$. In this regime, the piston rapidly adjusts so that $P_\mathrm{R}(t)\simeq P_\mathrm{ext}$, and the subsequent dynamics is dominated by a quasi--stationary drift  driven by the externally imposed pressure.

We can rationalize both the emergence and the magnitude of this velocity plateau within DDFT by integrating the particle current across the slit (see Eq.~\eqref{eq:v_material}), as shown in Appendix B. As a result, to leading order the plateau velocity is governed by the boundary pressure difference and is well approximated by
\begin{equation}
\langle v(t)\rangle_\mathrm{plateau}
\simeq
-\frac{A D}{N}\big(P_\mathrm{ext}-P_0\big).
\label{eq:plateau_v}
\end{equation}
Inserting the corresponding values of the pressures and surface concentration into this analytical estimate yields $\langle v^*(t) \rangle\simeq -0.929$, in excellent agreement with the numerical results.

\subsection{Non-equilibrium thermodynamics: entropy production, injected power and bath entropy}
\label{subsec:thermo_noneq}

\begin{figure}[ht!]
	\centering
	\includegraphics[width=1\linewidth]{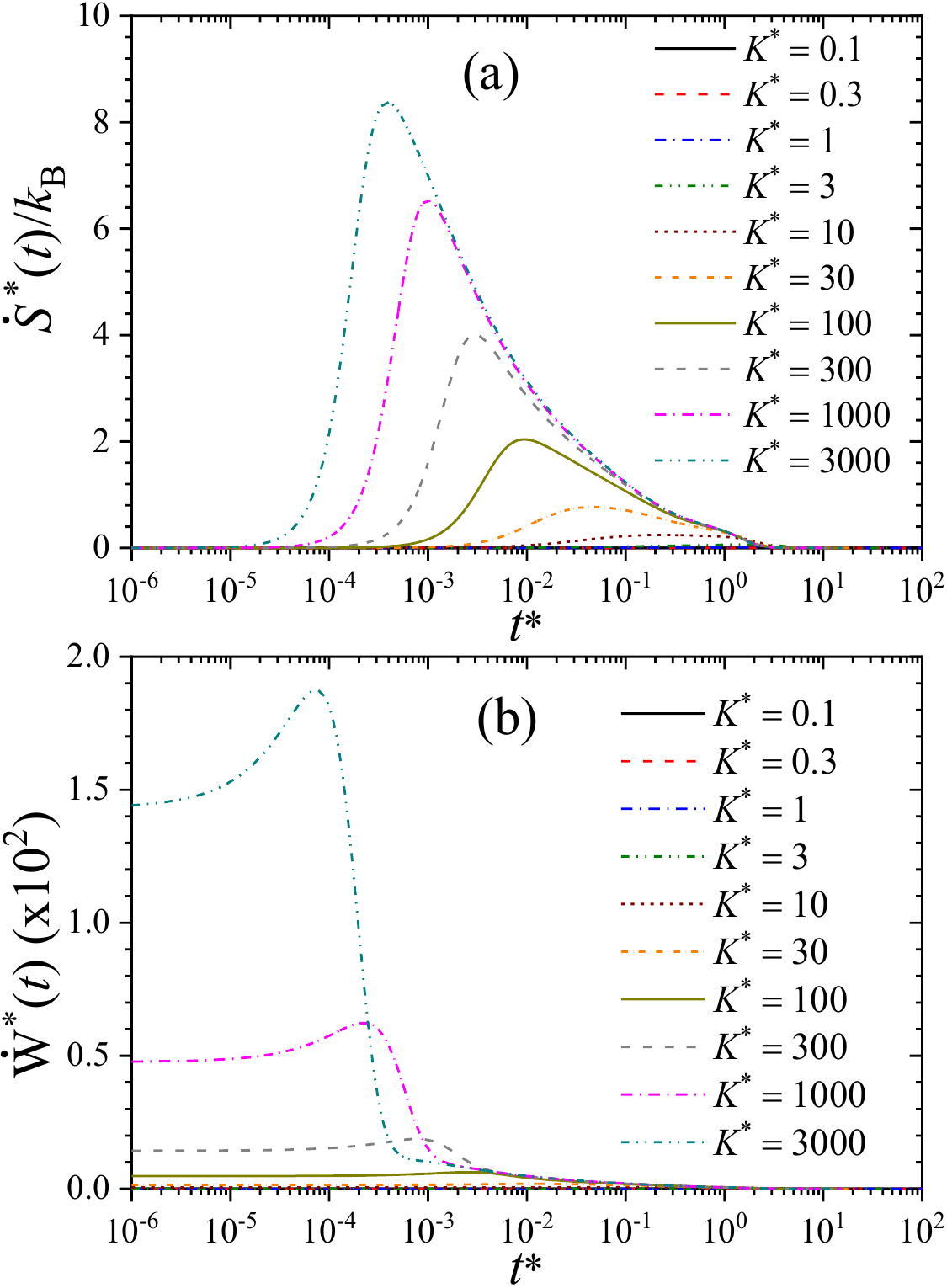}
	\caption{(a) Time evolution of the entropy production rate per unit of area for values of $K^*$ ranging from $0.1$ to $3000$. (b) Time evolution of the mechanical power per unit of area supplied by the compressing mobile piston to the hard-sphere fluid, for the same range of $K^*$ values.
}
\label{fig:Fig6}
\end{figure}

We now characterize the non-equilibrium thermodynamics of the compression process by examining the entropy production rate and the mechanical power injected by the piston (both per unit area) as functions of time for piston mobilities in the range $0.1 \le K^* \le 3000$. Fig~\ref{fig:Fig6}(a) shows the total reduced entropy production rate $\dot{S}^*(t)/k_\mathrm{B}$, which quantifies the heat irreversibly dissipated into the thermal reservoir due to the diffusive currents generated by the piston-driven compression. As expected, $\dot{S}^*(t)$ vanishes at the initial and final equilibrium states, and exhibits a pronounced maximum at an intermediate time, identifying the instant of largest dissipation. Increasing the piston mobility leads to a systematic increase of the peak amplitude and shifts its position to earlier times, reflecting the stronger and faster driving.

Interestingly, at sufficiently long times the decay of $\dot{S}^*(t)$ follows an almost common envelope for the different values of $K^*$. This behavior can be rationalized by noting that, once the piston has largely relaxed and the system enters the late-stage approach to equilibrium, the dynamics becomes controlled by the intrinsic diffusive relaxation of the confined fluid rather than by the piston mobility. In this regime the remaining density perturbations relax through similar slow diffusive modes, leading to a $K^*$--independent functional form of the long-time tail, while $K^*$ primarily affects the early-time buildup and the time at which the maximum dissipation occurs.

Fig.~\ref{fig:Fig6}(b) shows the corresponding reduced power injected by the piston, $\dot{W}^*(t)$, which is significantly larger in magnitude. In contrast to $\dot{S}^*(t)$, the injected power is nonzero immediately after the pressure jump, starting from the finite initial value $\dot{W}^*(0)=K^* P_0^*(P_\mathrm{ext}^*-P_0^*)$, and eventually decays to zero at long times as equilibrium is approached. Similar to the entropy production rate, $\dot{W}^*(t)$ exhibits a maximum at an intermediate time, whose position shifts to earlier times as $K^*$ increases. However, the post-maximum decay of $\dot{W}^*(t)$ does not collapse onto a common envelope and typically decreases more rapidly than $\dot{S}^*(t)$, consistent with the fact that the mechanical driving exerted by the piston becomes rapidly ineffective once the piston has adjusted, whereas dissipative relaxation persists over longer times due to diffusive redistribution within the fluid.
\begin{figure}[ht!]
	\centering
	\includegraphics[width=1\linewidth]{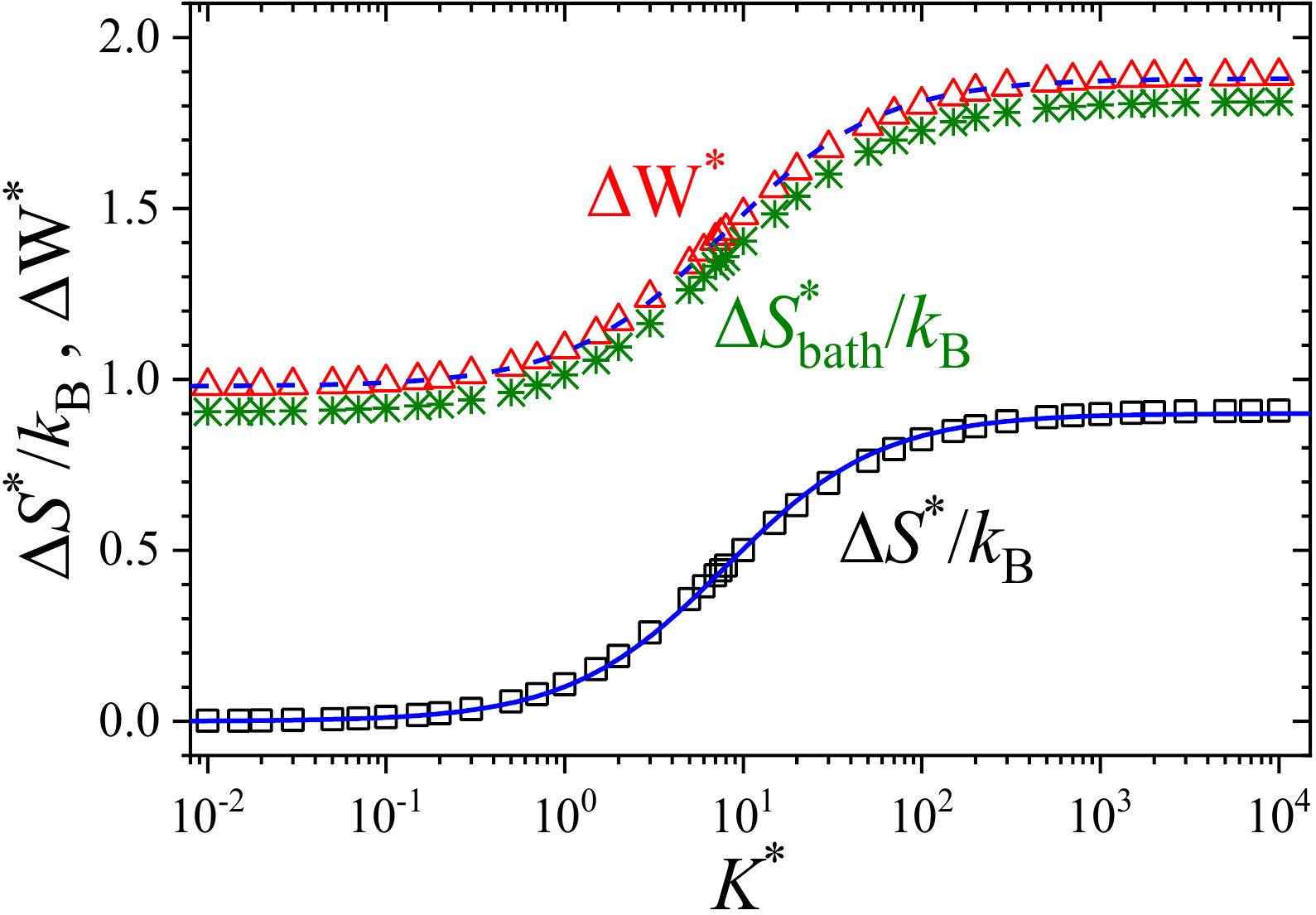}
	\caption{Total entropy production ($\Delta S^*$) and mechanical work supplied to the colloidal fluid ($\Delta W^*$) as functions of $K^*$.  Note that $-\Delta S^*/k_\mathrm{B} + \Delta W^*$ is a constant equal to the Helmholtz free-energy difference between the initial and final equilibrium states, $\Delta \mathcal{F}_\mathrm{eq}^*$. Green stars represent the variation of entropy of the thermal bath coupled to the system, $\Delta S_\mathrm{bath}^*=\Delta W^* - \Delta U_\mathrm{ext}^*$. Solid, dashed  and dotted blue lines correspond to fits according to Eqs.~\eqref{eq:ajusteDS}.
}
\label{fig:Fig7}
\end{figure}

Integrating the rates over time (see Eqs.~\eqref{eq:DS} and \eqref{eq:PRL}) yields the total dissipated heat and the total injected work per unit of area during the entire compression process. Fig.~\ref{fig:Fig7} shows $\Delta S^*/k_\mathrm{B}$ (black squares) and $\Delta W^*$ (red triangles) as functions of $K^*$. Both quantities display the same $K^*$--dependence, with $\Delta W^*$ shifted upward by a constant amount equal to the equilibrium Helmholtz free energy difference between the final and initial equilibrium states, i.e., $\Delta W^* - \Delta S^*/k_\mathrm{B} = \Delta \mathcal{F}_\mathrm{eq}^* = 0.980$, as required by the exact identity in Eq.~\eqref{eq:integrated_balance}.

In the quasi-static limit $K^* \to 0$, one finds $\Delta W^* \to \Delta \mathcal{F}_\mathrm{eq}^*$ and $\Delta S^*/k_\mathrm{B} \to 0$, indicating that the work supplied by the piston is fully converted into an increase of Helmholtz free energy without heat dissipation. As $K^*$ increases, both $\Delta W^*$ and $\Delta S^*/k_\mathrm{B}$ grow, first slowly, then more rapidly, and finally saturate towards their upper bound at large $K^*$. This saturation implies that neither the injected work nor the dissipated heat can increase without bound with piston mobility, consistent with the fact that the underlying diffusive currents are ultimately limited by the intrinsic transport properties of the colloidal fluid.

In addition, the entropy change of the thermal bath can be obtained directly from Eq.~\eqref{eq:Sbath}. The resulting values of $\Delta S_\mathrm{bath}^*$ are shown as green stars in Fig.~\ref{fig:Fig7}. As expected, $\Delta S_\mathrm{bath}^*$ follows a trend very similar to those of $\Delta S^*$ and $\Delta W^*$, differing only by a vertical offset. For $K^*\to 0$, the entropy production vanishes and the entropy variation of the bath exactly compensates that of the colloidal system, $\Delta S_\mathrm{bath}^* = -\,\Delta S_\mathrm{sys}^*$, corresponding to the minimum possible bath entropy change compatible with reversibility. In this limit, the bath merely exchanges heat with the system in order to maintain constant temperature, without any irreversible dissipation.

As $K^*$ increases, irreversible diffusive currents develop and the entropy production becomes finite. The thermal bath must then absorb not only the reversible heat associated with the configurational ordering of the fluid, but also the additional heat generated by irreversible dissipation. As a result, $\Delta S_\mathrm{bath}^*$ increases monotonically with $K^*$ and eventually saturates in the limit $K^*\to\infty$, where the entropy production (and thus the entropy transfer to the bath) reaches its maximal value.

The data are accurately described by the simple crossover form
\begin{equation}
    \Delta S^*/k_\mathrm{B} = S_\infty^*\bigg(1-\frac{1}{1+K^*/K_\mathrm{c}^*}\bigg),
    \label{eq:ajusteDS}
\end{equation}    
from which the corresponding work and bath--entropy variations follow as  $\Delta W^* = \Delta \mathcal{F}_\mathrm{eq}^* + \Delta S^*/k_\mathrm{B}$ and $\Delta S_\mathrm{bath}^*/k_\mathrm{B}=\Delta W^* - \Delta U_\mathrm{ext}^*$. The best-fit parameters are $S_\infty^* = 0.901 \pm 0.002$ and $K_\mathrm{c}^* = 7.9\pm 0.1$, shown as solid, dashed and dotted blue lines in Fig.~\ref{fig:Fig7}, respectively.

\begin{figure}[ht!]
	\centering
	\includegraphics[width=1\linewidth]{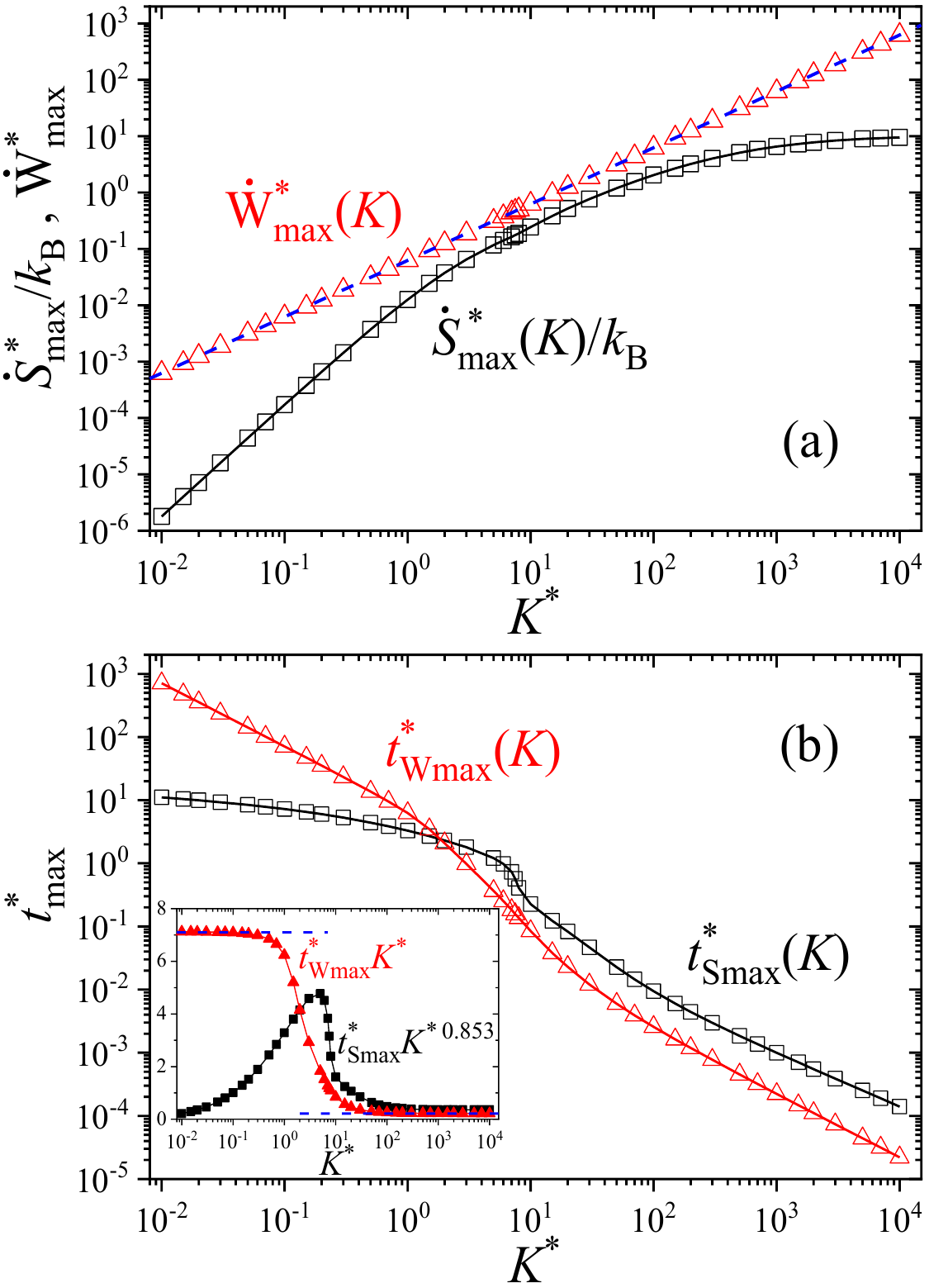}
	\caption{(a) Maximum value of the entropy production rate ($\dot{S}_{\mathrm{max}}^*$) and the power supplied by the mobile piston ($\dot{W}_{\mathrm{max}}^*$) as functions of $K$. Dashed blue line represents the theoretical prediction given by Eq.~\eqref{eq:Wmax}. (b) Corresponding times at which the maxima occur, $t_{\mathrm{Smax}}^*$ and $t_{\mathrm{Wmax}}^*$, as functions of $K$. Horizontal dashed blue lines are the limiting trends given by Eqs.~\eqref{eq:tW1} and \eqref{eq:tW2}.
Inset: $t_{\mathrm{Wmax}}^*\,K^*$ and $t_{\mathrm{Smax}}^*\,(K^*)^{0.853}$ versus $K$.}
	\label{fig:Fig8}
\end{figure}

We further characterize the non-equilibrium thermodynamics by analyzing both the heights and the temporal positions of the maxima of the injected power and the entropy production rate as functions of the piston mobility $K^*$. Fig.~\ref{fig:Fig8}(a) shows the peak values $\dot{W}_\mathrm{max}^*$ and $\dot{S}_\mathrm{max}^*/k_\mathrm{B}$ on double--logarithmic axes, while Fig.~\ref{fig:Fig8}(b) reports the times at which these maxima are reached, $t_{\mathrm{Wmax}}^*$ and $t_{\mathrm{Smax}}^*$.

As observed in Fig.~\ref{fig:Fig8}(a), the maximum injected power grows linearly with piston mobility, $\dot{W}_\mathrm{max}^* \propto K^*$, whereas the maximum entropy production rate follows a markedly different trend. For small mobilities, $\dot{S}_\mathrm{max}^*/k_\mathrm{B}$ increases approximately quadratically, $\dot{S}_\mathrm{max}^*/k_\mathrm{B}\sim (K^*)^2$, before crossing over to a saturation plateau at large $K^*$. Thus, while the height of the power peak continues to increase with $K^*$, the peak entropy production becomes bounded. This boundedness can be understood from the quadratic dissipation functional underlying the DDFT dynamics. Indeed, the entropy production rate reads Eq.~(\ref{eq:entropyproduction}),
$\dot S(t)\propto \int dz\, J(z,t)^2/[D\,\rho(z,t)]$, so that increasing $\dot S$ requires a sustained increase of the diffusive current. In the high-mobility regime, $K^*\gg 1$, the piston rapidly adjusts such that $P_R(t)\simeq P_\mathrm{ext}$, and the subsequent relaxation is controlled by transport inside the slit rather than by the piston mobility (cf.\ the nearly $K^*$-independent long-time tails of $\dot S^*(t)$ in Fig.~\ref{fig:Fig6}(a)). In this regime the net center-of-mass velocity approaches the plateau value, Eq.~(\ref{eq:plateau_v}), which is essentially $K^*$-independent; a scaling estimate $J\simeq \rho\langle v\rangle_\mathrm{plateau}$ then yields $\dot S_\mathrm{max}\sim A\int dz\,\rho\,\langle v\rangle_\mathrm{plateau}^2/D$, rationalizing a diffusion-limited upper bound for $\dot{S}_\mathrm{max}^*$ as $K^*\to\infty$. 
For all values of $K^*$ one finds $\dot{W}_\mathrm{max}^*>\dot{S}_\mathrm{max}^*/k_\mathrm{B}$, with both quantities becoming comparable only in the intermediate crossover regime $K^*\sim 5$--$10$.

The linear scaling of $\dot{W}_\mathrm{max}^*$ can be understood directly from the expression for the mechanical power injected by the piston. Using Eq.~\eqref{eq:mechpower_def} and rewriting the power in terms of the instantaneous pressure exerted by the colloids on the piston, $P_\mathrm{R}^*(t)$, one obtains $\dot{W}^*=-P_\mathrm{R}^*\,\dot L^*=K^*\,P_\mathrm{R}^*\big(P_\mathrm{ext}^*-P_\mathrm{R}^*\big)$, which is a concave parabola as a function of $P_\mathrm{R}^*$. The maximum of this parabola is attained when $d\dot{W}^*/dP_\mathrm{R}^*=K^*(P_\mathrm{ext}^*-2P_\mathrm{R}^*)=0$, i.e., at $P_\mathrm{R}^*=P_\mathrm{ext}^*/2$, independently of $K^*$. Substituting this value yields 
\begin{equation}
\dot{W}_\mathrm{max}^*=(P_\mathrm{ext}^{*2}/4)\,K^*=0.0625\,K^*,
\label{eq:Wmax}
\end{equation}
as shown by the blue dashed line in Fig.~\ref{fig:Fig8}(a), in excellent agreement with the numerical data.

The temporal positions of the maxima are summarized in Fig.~\ref{fig:Fig8}(b). For the injected power, the peak time $t_{\mathrm{Wmax}}^*$ exhibits two asymptotic regimes at low and high mobilities, in both of which it decreases as $1/K^*$ but with markedly different prefactors, separated by a crossover at intermediate $K^*$. In the low--mobility regime ($K^*\ll 1$), the analytical expression for $P_\mathrm{R}(t)$ in terms of the rescaled time $x=t/\tau_K$ can be used, and imposing the condition $P_\mathrm{R}^*=P_\mathrm{ext}^*/2$ yields the numerical solution $x_\mathrm{max}=t_{\mathrm{Wmax}}^*/\tau_K^*=0.68$. Using the definition of $\tau_K^*$ (see Eq.~\eqref{eq:tauK}) then gives
\begin{equation}
t_{\mathrm{Wmax}}^*=7.03/K^* \ \ \ \ (K^* \ll 1).
\label{eq:tW1}
\end{equation}

Repeating the same procedure in the opposite limit of large mobility ($K^*\gg 1$), using the corresponding high--$K^*$ scaling form, yields $t_{\mathrm{Wmax}}^*/\tau_K^{\prime *}=0.561$ and therefore 
\begin{equation}
t_{\mathrm{Wmax}}^*= 0.216/K^* \ \ \ \ (K^*\gg 1).
\label{eq:tW2}
\end{equation}

This behavior is clearly confirmed in the inset of Fig.~\ref{fig:Fig8}(b), where $t_{\mathrm{Wmax}}^*K^*$ is plotted as a function of $K^*$ and forms two well-defined plateaus at $7.03$ and $0.216$.

A particularly revealing feature of Fig.~\ref{fig:Fig8}(b) is the crossing between $t_{\mathrm{Wmax}}^*$ and $t_{\mathrm{Smax}}^*$ occuring at $K^*=2$. For large mobilities, the piston injects mechanical power very rapidly, while the dissipation associated with diffusive relaxation occurs at later times, leading to $t_{\mathrm{Wmax}}^*<t_{\mathrm{Smax}}^*$. In contrast, for small mobilities the piston advances slowly and the fluid adapts diffusively on shorter timescales, generating dissipative currents already while the injected power is still building up, so that $t_{\mathrm{Wmax}}^*>t_{\mathrm{Smax}}^*$. This crossing therefore provides a clear dynamical signature of the competition between mechanical driving and diffusive relaxation.

The inset of Fig.~\ref{fig:Fig8}(b) also shows that at large $K^*$ the entropy--production peak time follows a nontrivial power law, $t_{\mathrm{Smax}}^*\sim (K^*)^{-0.853}$, reflecting the fact that the location of the dissipation maximum is controlled by the intrinsic diffusive relaxation of the confined fluid, rather than directly by the piston mobility.

\begin{figure}[ht!]
	\centering
	\includegraphics[width=1\linewidth]{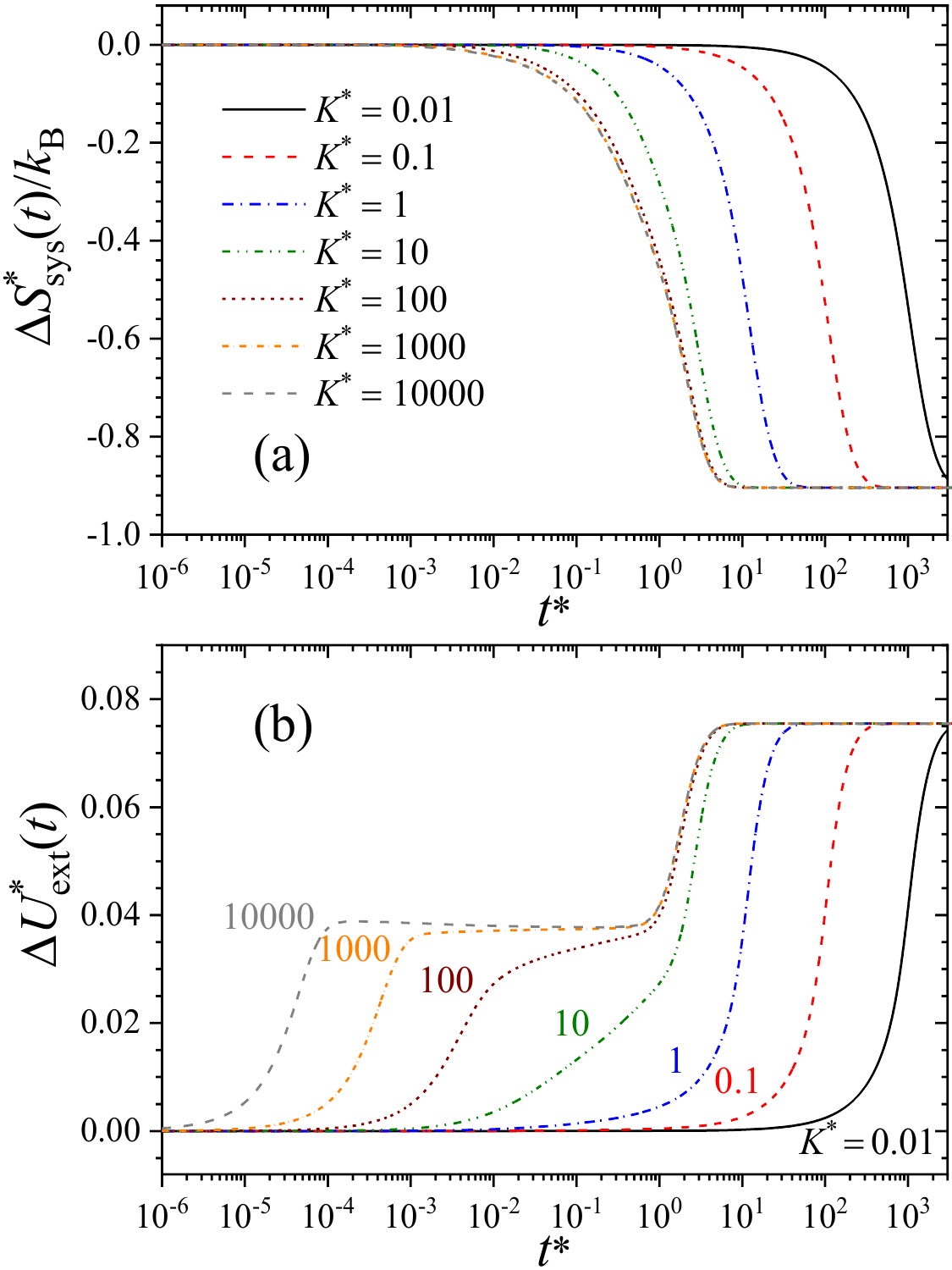}
	\caption{(a) Temporal evolution of the intrinsic (configurational) entropy variation of the colloidal system, $\Delta S_\mathrm{sys}^*(t)$, for piston mobilities $0.01 \le K^* \le 10000$. (b) Corresponding evolution of the external confinement potential energy change, $\Delta U_\mathrm{ext}^*(t)$.
}
\label{fig:Fig9}
\end{figure}

Finally, in order to complete the thermodynamic characterization of the compression process, we analyze the full time evolution of the configurational entropy change of the colloidal system, $\Delta S_\mathrm{sys}(t)$, and of the corresponding change in external potential energy, $\Delta U_\mathrm{ext}(t)$, both measured with respect to their initial equilibrium values. These are genuine non-equilibrium dynamical quantities which, in the long-time limit $t \to \infty$, converge to their final equilibrium values. By construction, they are related to the instantaneous free energy of the system through $- T \Delta S_\mathrm{sys}(t) + \Delta U_\mathrm{ext}(t) 
= \Delta \mathcal{F}(t)$.

Figs.~\ref{fig:Fig9}(a) and \ref{fig:Fig9}(b) display respectively the temporal evolution of $\Delta S_\mathrm{sys}(t)$ and $\Delta U_\mathrm{ext}(t)$ per unit area, in reduced units, for piston mobilities $0.01 \le K^* \le 10000$. As expected, the system entropy decreases as compression proceeds, reflecting the progressive reduction of accessible configurations as the piston advances. For $K^*=0.01$, the entropy closely follows quasi-equilibrium values corresponding to the instantaneous piston position. As $K^*$ increases, this quasi-static behavior progressively breaks down and the entropy decrease becomes steeper. In the high-mobility regime, a saturation behavior emerges again: for sufficiently large $K^*$ the curves $\Delta S_\mathrm{sys}(t)$ converge towards a limiting profile, indicating that the fluid cannot sustain arbitrarily large internal currents and that the dynamics becomes diffusion-limited.

The behavior of $\Delta U_\mathrm{ext}(t)$ is more subtle. For $K^*=0.01$, the external potential energy increases smoothly through quasi-equilibrium states, with the process accelerating as $K^*$ increases. However, for $K^* \gtrsim 10$ an intermediate regime becomes visible, which evolves into a pronounced two-step behavior for $K^* \gtrsim 1000$. In this regime, $\Delta U_\mathrm{ext}(t)$ develops a transient pseudo-plateau, indicating that the fluid continues to be confined while its external potential energy remains nearly constant. Remarkably, this time window coincides with the plateau previously observed in the center-of-mass velocity $\langle v^*(t)\rangle$.

A closer inspection reveals that $\Delta U_\mathrm{ext}(t)$ may even exhibit a weak non-monotonic behavior within this intermediate regime: despite the ongoing confinement, the external potential energy can temporarily decrease before rising again towards its final equilibrium value. This reflects a transient structural reorganization of the density profile. After the rapid initial compression, particles accumulate near the piston, increasing $U_\mathrm{ext}$. During the subsequent diffusive relaxation, the density redistributes more uniformly across the slit, temporarily reducing the average contribution of the external field before the final equilibrium layering structure is established. For $t^* \gtrsim 0.1$, the second increase sets because the density perturbation reaches the left wall and the system approaches its final compressed equilibrium state.

This non-monotonicity provides a direct signature of the competition between advective confinement imposed by the piston and diffusive structural reorganization of the colloidal fluid. It highlights that thermodynamic quantities in driven confined systems cannot, in general, be inferred solely from the instantaneous piston position, but depend on the internal non-equilibrium density profile. In this sense, the transient behavior of $\Delta U_\mathrm{ext}(t)$ offers a clear macroscopic manifestation of the underlying microscopic relaxation dynamics.

\section{Conclusions}
\label{sec:conclusions}

In this work we have investigated the non-equilibrium response of a confined colloidal fluid driven by a mobile boundary under an externally imposed pressure jump. Within a DDFT framework coupled to an overdamped piston equation of motion, we have followed the full relaxation process from an initial equilibrium state to a compressed equilibrium configuration. By systematically varying the piston mobility $K^*$ over several orders of magnitude, we identified distinct dynamical regimes governing the compression dynamics.

The relaxation process is controlled by the competition between externally imposed mechanical driving and intrinsic diffusive relaxation of the confined fluid. As the piston mobility increases, the system crosses over from a quasi–static regime to a diffusion–limited regime in which the piston dynamics becomes effectively slaved to the internal relaxation of the fluid.

At the microscopic level, the density and current fields reveal how advective piston motion competes with diffusive particle transport. The present approach also makes it possible to resolve the spatial asymmetries generated by the moving boundary, which arise because the propagation of pressure through the confined fluid is not instantaneous but limited by diffusive transport. For small mobilities ($K^*\ll1$), the system remains close to local equilibrium and the compression proceeds quasi–statically. In contrast, for large mobilities ($K^*\gg1$), the piston responds rapidly to the pressure jump, producing strong non–equilibrium density and current asymmetries. In this regime the piston trajectory $L(t)$ and the pressure–position relation $P_\mathrm{R}(L)$, converge toward universal $K^*$–independent forms.

From a macroscopic perspective we analyzed the evolution of piston position, pressure response, injected mechanical power, and entropy production. Both the total injected work and the entropy production remain bounded, with the work ranging from the quasi–static limit $\Delta W^* \to \Delta \mathcal{F}_\mathrm{eq}^*$ to a finite maximum in the high–mobility limit. These bounds reflect fundamental constraints imposed by diffusive transport in the confined fluid. Further insight is provided by the time evolution of the configurational entropy change $\Delta S_\mathrm{sys}(t)$ and the external potential energy $\Delta U_\mathrm{ext}(t)$, which satisfy $-T\Delta S_\mathrm{sys}(t)+\Delta U_\mathrm{ext}(t)=\Delta\mathcal{F}(t)$. While $\Delta S_\mathrm{sys}(t)$ decreases monotonically due to progressive confinement, $\Delta U_\mathrm{ext}(t)$ can display transient non–monotonic behavior in the high–mobility regime, reflecting the decoupling between rapid geometric confinement and slower diffusive particle redistribution.

Although the present study focuses on hard–sphere colloids, the thermodynamic structure identified here—bounded work and entropy production together with a mobility–controlled dynamical crossover—is not specific to the interaction potential and reflects generic features of overdamped driven systems.

The present results demonstrate that mechanically driven confinement provides a minimal and controlled setting to investigate non-equilibrium thermodynamics with self-consistent boundary dynamics, a situation of broad relevance in soft matter and statistical physics. Future work could explore memory effects in compression and expansion protocols~\cite{C9SM02005E,GrohDzubiella2025Kovacs}, as well as colloidal systems with responsive particle sizes~\cite{moncho-jorda2023,moncho-jorda2024external,lopez-molina2024DFTResponsiveHS,lopez-molina2024nonequilibrium,piston_karg1,piston_karg2,Isa_compression}. A further extension would be to incorporate interaction–induced dissipation within the Power Functional Theory framework~\cite{schmidt2013,Schmidt2022}, which allows the inclusion of memory effects and nonlocal dynamical correlations beyond standard DDFT.

\acknowledgments

A.M.J. and J.L.M. thanks grant no. PID2022-136540NB-I00 awarded by MICIU/AEI/10.13039/501100011033 and ERDF, a way of making Europe, and Project No. A-EXP-359-UGR23, co-funded by Junta de Andalucía-Consejería de Universidad, Investigación e Innovación and by the European Union under the FEDER Andalucía 2021–2027 program J.D. acknowledges funding by the Deutsche Forschungsgemeinschaft (DFG) via the Research Unit FOR 5099 “Reducing complexity of nonequilibrium systems” (Project No. 431945604). The authors thank Servicio de Supercomputación UGR and PROTEUS, the supercomputing center of the Institute Carlos I for Theoretical and Computational Physics, for the computational time provided. We thank Dr. Sebastien Groh (Univ. Freiburg, Germany) for helpful discussions.

 \appendix

 \section{Non-equilibrium currents}

Fig.~\ref{fig:Fig10} displays the local particle currents $J^*(z,t)$ generated by the piston motion during the compression process, as a function of time, for the same representative piston mobilities $K^* = 0.1$, $10$, and $1000$. At both the initial and final times, the currents vanish, as the system is in equilibrium. Once the compression protocol is activated, nonzero currents develop during the early stages of the process and subsequently decay progressively to zero as the system relaxes toward the final equilibrium state. The negative sign of the currents simply reflects that particle velocities are directed along the negative $z$--direction as a result of the compression.
\begin{figure}[ht!]
	\centering
	\includegraphics[width=1\linewidth]{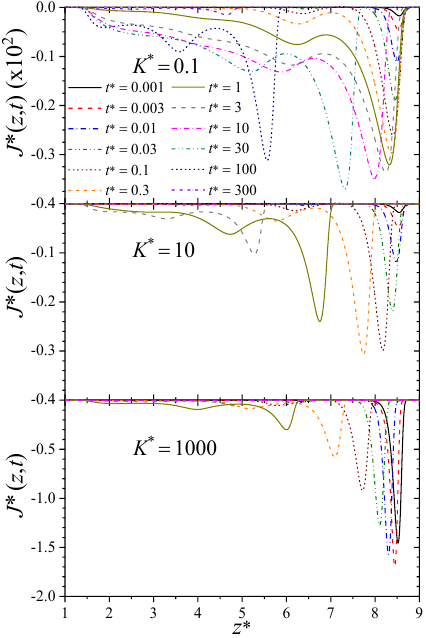}
	\caption{Time evolution of the non-equilibrium particle fluxes induced by the mobile wall for $K^* = 0.1$, $10$, and $1000$. Note that each panel is displayed using a different vertical scale.
}
\label{fig:Fig10}
\end{figure}

In all cases, the currents are distributed inhomogeneously throughout the system. Their spatial structure closely follows that of the density profiles shown in Fig.~\ref{fig:Fig2}(b), with local maxima approximately coinciding with the density peaks. In addition, the largest current amplitudes are consistently observed in the region adjacent to the piston, where the perturbation induced by the piston motion is strongest. Moving away from the piston toward the fixed wall, the magnitude of the current peaks progressively decreases, indicating the gradual transmission and relaxation of the perturbation across the confined fluid.

For the slow piston mobility $K^* = 0.1$, the currents remain very small at all times, confirming that the system stays close to equilibrium throughout the compression. Increasing the piston mobility to $K^* = 10$ leads to an approximately one order of magnitude increase in the current amplitudes, while the maximum currents are reached at earlier stages of the compression. In the case of a very fast piston, $K^* = 1000$, the currents attain their maximum values at very short times, of the order of $t^* \simeq 3 \times 10^{-3}$, and subsequently decay in time. Although the piston mobility is increased by two orders of magnitude compared to the $K^* = 10$ case, the corresponding increase in the current amplitudes is limited to about a factor of two, indicating that the currents are attaining a saturation behavior for $K\rightarrow \infty$. In other words, the fluid velocity cannot be further increased by enhancing the piston mobility, being bounded by the intrinsic diffusive timescale of the colloids.

\section{Estimate of plateau velocity}
 \label{sec:appendixA}

The total current can be decomposed into ideal, excess, and external--field contributions, namely
\begin{equation}
J(z,t)=-D\left[
\frac{\partial \rho}{\partial z}+\rho\,\frac{\partial \mu_\mathrm{ex}}{\partial z}+\rho\left(\frac{\partial V_\mathrm{L}}{\partial z}+\frac{\partial V_\mathrm{R}}{\partial z}\right)\right],
\end{equation}
where $\mu_\mathrm{ex}(z,t)=\delta F_\mathrm{ex}/\delta \rho$ is the excess chemical potential.

Integrating the first (ideal) contribution over the accessible domain yields
\begin{equation}
\int_R^{L_0-R}dz\,\frac{\partial \rho(z,t)}{\partial z}
=
\rho(L_0-R,t)-\rho(R,t)
\simeq 0,
\end{equation}
which can be safely ignored since the repulsive wall--particle interactions strongly suppress the colloid density at contact, effectively excluding particles from these regions.

Analogously, integrating the external--field contribution over $z$ and using Eqs.~\eqref{eq:PL} and \eqref{eq:PR} leads to the difference of the instantaneous wall pressures,
\begin{equation}
\int_R^{L_0-R}dz\,\rho
\left(
\frac{\partial V_\mathrm{L}}{\partial z}
+
\frac{\partial V_\mathrm{R}}{\partial z}
\right)
=
P_\mathrm{R}(t)-P_\mathrm{L}(t)
\simeq
P_\mathrm{ext}-P_0,
\end{equation}
where, in this intermediate time window at large $K^*$, the pressure on the left wall remains close to its initial value, $P_\mathrm{L}(t)\simeq P_0$, while the pressure on the piston matches the imposed external pressure, $P_\mathrm{R}(t)\simeq P_\mathrm{ext}$.

Finally, although the excess contribution
$I_\mathrm{ex}(t)=-D\int_R^{L_0-R}dz\,\rho(z,t)\,\partial_z\mu_\mathrm{ex}(z,t)$
cannot be evaluated analytically, direct numerical calculations show that $I_\mathrm{ex}(t)$ remains much smaller than the imposed pressure imbalance, $|I_\mathrm{ex}(t)|\ll |P_\mathrm{R}(t)-P_\mathrm{L}(t)|$, throughout the plateau regime for $K^*\gg 1$. Physically, $\partial_z\mu_\mathrm{ex}$ is tied to the oscillatory layered microstructure of the hard--sphere fluid and changes sign across the slit, producing strong spatial cancellations upon integration. As a result, interactions strongly affect the local structure and currents, but contribute only weakly to the net center--of--mass transport. Hence, to leading order the plateau velocity is given by eq.~(\ref{eq:plateau_v}).

\bibliographystyle{apsrev4-2}
\bibliography{DDFT_HS}

\end{document}